\documentclass[journal]{IEEEtran}

\usepackage[binary-units=true]{siunitx}
\usepackage{xcolor}
\usepackage{tikz}
\usepackage[mode=buildnew]{standalone}
\usepackage{enumerate} 
\usepackage{siunitx}
\usepackage{amssymb}
\usepackage{amsmath}
\usepackage{booktabs}
\usepackage{tabularx}
\usepackage{makecell}
\usepackage{multirow}
\usepackage{rotating}
\usepackage{hyperref}

\usepackage{listings}
\definecolor{codeblue}{rgb}{0.00,0.20,0.60}
\definecolor{codegreen}{rgb}{0.25,0.50,0.35}
\definecolor{codepurple}{rgb}{0.50,0.00,0.50}
\definecolor{white}{rgb}{1.0,1.0,1.0}
\definecolor{codebg}{rgb}{0.97,0.97,0.97}

\lstdefinestyle{mystyle}{
  language=Python,
  basicstyle=\footnotesize\ttfamily,
  keywordstyle=\color{codeblue}\bfseries,
  commentstyle=\color{codegreen}\itshape,
  stringstyle=\color{codepurple},
  numberstyle=\tiny\color{codegray},
  showstringspaces=false,
  keepspaces=true,
  columns=fullflexible,
  breaklines=true,
  frame=single,
  rulecolor=\color{white},
  framesep=2mm,
  escapeinside={|}{|},
  mathescape=true
}

\usepackage{academicons}
\usepackage{xcolor}
\newcommand{\orcid}[1]{\href{https://orcid.org/#1}{\textcolor[HTML]{A6CE39}{\aiOrcid}}}

\usepackage{comment}
\DeclareSIUnit{\nothing}{\relax}

\title{Improving Autonomous Nano-drones Performance via Automated End-to-End Optimization and Deployment of DNNs}

\author{\IEEEauthorblockN{
Vlad Niculescu\IEEEauthorrefmark{1}, %\orcid{},
Lorenzo Lamberti\IEEEauthorrefmark{2}, % \orcid{0000-0003-1659-618X},
Francesco Conti\IEEEauthorrefmark{2}, %\orcid{0000-0002-7924-933X},
Luca Benini\IEEEauthorrefmark{1}\IEEEauthorrefmark{2}, %\orcid{0000-0001-8068-3806},
and Daniele Palossi\IEEEauthorrefmark{1}\IEEEauthorrefmark{3}}% \orcid{0000-0003-4487-0836}

\IEEEauthorblockA{\IEEEauthorrefmark{1}Integrated Systems Laboratory - ETH Z\"urich, Switzerland}

\IEEEauthorblockA{\IEEEauthorrefmark{2} Department of Electrical, Electronic and Information Engineering - University of Bologna, Italy}

\IEEEauthorblockA{\IEEEauthorrefmark{3} Dalle Molle Institute for Artificial Intelligence - USI-SUPSI, Switzerland}

vladn@iis.ee.ethz.ch, lorenzo.lamberti@unibo.it, f.conti@unibo.it, luca.benini@iis.ee.ethz.ch, daniele.palossi@idsia.ch
}

\definecolor{somegray}{rgb}{0.5, 0.5, 0.5}
\newcommand{\darkgrayed}[1]{\textcolor{somegray}{#1}}
\makeatletter
\newcommand*\titleheader[1]{\gdef\@titleheader{#1}}
\AtBeginDocument{%
  \let\st@red@title\@title
  \def\@title{%
    \vskip-1.7em
    \bgroup\normalfont\large\centering\@titleheader\par\egroup
    \vskip0.4em\st@red@title}
}

\makeatother
\titleheader{\darkgrayed{This paper has been accepted for publication in the IEEE Journal on Emerging and Selected Topics in Circuits and Systems (JETCAS) \copyright 2021 IEEE.}}
\begin{document}
\maketitle

\begingroup\renewcommand\thefootnote{\textsection}
\endgroup

\begin{abstract}
The evolution of energy-efficient ultra-low-power (ULP) parallel processors and the diffusion of convolutional neural networks (CNNs) are fueling the advent of autonomous driving nano-sized unmanned aerial vehicles (UAVs).
These sub-\SI{10}{\centi\meter} robotic platforms are envisioned as next-generation ubiquitous smart-sensors and unobtrusive robotic-helpers.
However, the limited computational/memory resources available aboard nano-UAVs introduce the challenge of minimizing and optimizing vision-based CNNs -- which to date require error-prone, labor-intensive iterative development flows.
This work explores methodologies and software tools to streamline and automate all the deployment of vision-based CNN navigation on a ULP multicore system-on-chip acting as a mission computer on a Crazyflie 2.1 nano-UAV.
We focus on the deployment of PULP-Dronet~\cite{palossi2019IOTJ}, a state-of-the-art CNN for autonomous navigation of nano-UAVs, from the initial training to the final closed-loop evaluation.
Compared to the original hand-crafted CNN, our results show a $2\times$ reduction of memory footprint and a speedup of $1.6\times$ in inference time while guaranteeing the same prediction accuracy and significantly improving the behavior in the field, achieving: \textit{i}) obstacle avoidance with a peak braking-speed of \SI{1.65}{\meter/\second} and improving the speed/braking-space ratio of the baseline, \textit{ii}) free flight in a familiar environment up to \SI{1.96}{\meter/\second} (\SI{0.5}{\meter/\second} for the baseline), and \textit{iii}) lane following on a path featuring a \SI{90}{\deg} turn -- all while using for computation less than 1.6\% of the drone's power budget.
To foster new applications and future research, we open-source all the software design in a ready-to-run project compatible with the Crazyflie 2.1. 
\end{abstract}

\section*{Supplementary material}
Open-source code and dataset are available at: \url{https://github.com/pulp-platform/pulp-dronet}.
In-field experiments video footage at: \url{https://youtu.be/41IwjAXmFQ0}, \url{https://youtu.be/Cd9GyTl6tHI}.

\IEEEpeerreviewmaketitle

\bstctlcite{IEEEexample:BSTcontrol}

\section{Introduction}\label{sec:introduction}

\IEEEPARstart{I}{n} the past years, unmanned aerial vehicles (UAVs) have been adopted in a wide range of applications, such as surveillance and inspection of hazardous areas~\cite{palossi2019IOTJ, gomez2017precise}. 
Nano-size UAVs, with a form factor of a few centimeters and a weight of tens of grams, are the ideal candidates for fully autonomous indoor navigation as they can safely operate near humans and reach narrow spots with their reduced dimensions~\cite{palossi2019IOTJ,palossi2019DCOSS,palossi2021fully}.
However, these platforms have a total power envelope of a few Watts, of which only $5-15\%$ is allotted for computation, making it challenging to deploy real-time navigation pipelines directly onboard~\cite{Wood2017}.
Furthermore, the small physical footprint and limited payload that can be carried by nano-UAVs constrains the battery and printed circuit board sizes.
Overall, these constraints mean that onboard computing devices need to have the physical footprint, power envelope, and on-chip memory of a typical microcontroller unit (MCU).

% Traditional approach - SLAM
For traditional UAVs, the classical approach for autonomous navigation is simultaneous-localization-and-mapping (SLAM), which creates a map of the environment and plans the trajectory according to it~\cite{loianno2018special}.
Classical SLAM is too computationally intensive to be feasible on nano-UAVs.
An alternative emerging approach is to infer relevant navigation information directly from onboard sensors and cameras, using machine learning-based algorithms.
In particular, deep convolutional neural networks (CNNs) have recently proved to provide good performance in autonomous navigation, at a fraction of the cost of SLAM: enough to run practical navigation tasks directly on highly resource-constrained platforms~\cite{palossi2019DCOSS,zhao2020learning}.
Still, achieving more sophisticated navigation skills requires to deploy more complex CNNs under even stricter real-time constraints, to promptly react to challenging dynamic environments, avoid collisions, plan new routes, etc.
Therefore, it is imperative to look for strategies to minimize the models' complexity and footprint while maintaining high accuracy.

Recently, low-power multi-core System-on-Chips (SoCs) have been introduced as potentially ideal devices to combine an MCU's flexibility with AI-oriented compute acceleration capabilities~\cite{gap8, MAX78000}.
At their peak performance, these devices deliver up to 10--100$\times$ better performance and efficiency than conventional MCUs, constituting an ideal platform for fully onboard DNN-driven autonomous navigation.
However, their complex architecture, together with the non-trivial requirements of DNN-based algorithms, requires a complex procedure including training, quantization, and a difficult hand-tuning phase to maximize performance on the final target -- a critical step to achieve high frame rate and thus good in-field navigation performance.

In this work, we focus on automating the end-to-end deployment of a DNN-based neural flight controller on top of a nano-UAV employing the GreenWaves Technologies (GWT) GAP8 SoC~\cite{gap8} -- one of the most advanced commercially available AI-oriented SoCs suitable to nano-size drones.
We evaluate two distinct toolsets available for GAP8, namely the GAP\textit{flow} provided by GWT and the open-source NEMO/DORY flow fostered by the research community~\cite{burrello2020dory}.
Specifically, we adapt and tune these flows to automatically deploy a CNN for autonomous navigation based on the state-of-the-art (SoA) PULP-Dronet~\cite{palossi2019IOTJ}.
PULP-Dronet is a residual network used to drive a nano-UAV through an interior (e.g., a corridor) or exterior (e.g., street) environment, deriving a \textit{probability of collision}, used for obstacle avoidance, and a \textit{steering angle} to keep within a lane -- implemented as a classification and a regression task, respectively.

Differently from the seminal PULP-Dronet, which relied on 16-bit fixed-point data representation, we focus on fully automated deployment, including network quantization to 8-bits, data tiling, code generation for the GAP8 SoC, evaluation of performance on the regression and classification tasks.
We also improve the integration of the new PULP-Dronet with the flight controller, with a more robust approach to deal with situations where the network's output is not providing strong guidance. 
We compare our results in terms of accuracy to the original PULP-Dronet showing that the prediction capability is maintained ($\sim$90\% for the classification despite the stronger quantization).
Our results show a throughput up to \SI{19}{frame/\second}, improved by a factor of up to 1.6$\times$ and total energy consumption of $\sim3-$\SI{4}{\milli \joule/frame}, which is $\sim44-58\%$ less than our baseline.

Moreover, we contribute a thorough exploration of the real-world performance of the CNN in exterior and interior environments, evaluating the drone's adherence to expected behavior in several controlled experiments performed in a room equipped with a Vicon motion capture system.
We individually assess the obstacle avoidance and steering capabilities.
We find that the drone can stop \SI{0.42}{\meter} away from a dynamic obstacle that appears \SI{1.5}{\meter} in front of the drone while flying with \SI{1.41}{\meter/\second}, with a significant 25.3\% improvement in the speed/braking-distance ratio vs. the original PULP-Dronet.
Furthermore, we also demonstrate the capability of the drone to fly an angled narrow tunnel, and we record the trajectories for various drone velocities.
We also evaluate the free-flight capabilities of the drone in a controlled indoor environment, achieving \SI{110}{\meter} path in \SI{56}{\second}, which marks an improvement of $\sim4\times$ vs. our baseline.

Our new, streamlined approach significantly improves the autonomous flight capabilities of PULP-Dronet while freeing up resources (i.e., reduced memory footprint and inference time), which allows the system to handle even more tasks (e.g., localization, detection, tracking, etc.).
Also, we investigate the generalization capabilities of the drone flying in new environments that are not captured by the training dataset.
Among all considered environments, we recorded the longest flight time of \SI{171}{\second} in an urban street.

\section{Related Work}\label{sec:related}

\begin{table}[t]
\caption{UAVs taxonomy by vehicle class-size~\cite{palossi2017ultra}.}
\begin{center}
\begin{tabularx}{\linewidth}{lccc}
\toprule
\small
Vehicle class & $\oslash$ : Weight [cm:kg] & Power [W] & Onboard Device\\
\midrule
\text{\textit{standard-size}~\cite{Smolyanskiy}} & $\sim$ 50 : $\geq$ 1	& $\geq$ 100 & Desktop\\
\text{\textit{micro-size}~\cite{weedNet}}	& $\sim$ 25 : $\sim$ 0.5 & $\sim$ 50 & Embedded\\
\text{\textit{nano-size}~\cite{palossi2019IOTJ}} & $\sim$ 10 : $\sim$ 0.01 & $\sim$ 5 & MCU\\
\text{\textit{pico-size}~\cite{Wood2017}} & $\sim$ 2 : $\leq$ 0.001 & $\sim$ 0.1 & ULP\\
\bottomrule
\end{tabularx}
\end{center}
\label{tab:taxonomy}
\end{table}

\subsubsection*{Standard/micro-sized UAVs}
%taxonomy
Customarily, we divide UAVs into four categories, shown in Table~\ref{tab:taxonomy}, according to size, weight, total power consumption, and onboard processing platform.
The latter two characteristics are directly linked, as the budget for onboard electronics is limited to $\sim$5-15\% of the total~\cite{Wood2017}.
The overwhelming majority of complex robotic perception algorithms have been demonstrated aboard standard- and micro-sized UAVs~\cite{Smolyanskiy,weedNet,Bodin}, which feature powerful onboard computers, often equipped with GPUs, such as NVIDIA Jetson TX1/TX2.
The sophisticated functionality that can be achieved with these platforms include onboard autonomous navigation in an unstructured natural environments~\cite{Smolyanskiy}; and search for particular objects in the field of view using semantic segmentation, for example for smart agriculture~\cite{weedNet}.
To work, these functionalities need multiple CNNs and other non-neural algorithms, such as visual SLAM, to enable multiple concurrent tasks, such as pose estimation, collision avoidance, and trajectory planning.

\subsubsection*{MCU-based nano-sized UAVs}
Highly miniaturized nano-size UAVs~\cite{palossi2019IOTJ,zhao2020learning,palossi2021fully} have a diameter of about \SI{10}{\centi\meter}, weigh only a few tens of grams, and have a total power budget of a few Watts -- not enough to directly support the functionality discussed above, with only simple MCU-class devices and a few \SI{}{\mega\byte} of memory onboard.
We can distinguish three categories of solutions to enable autonomous navigation despite these limitations: restricting to limited functionality to minimize the workload~\cite{zhao2020learning, lambert, shi_neural_swarm, Andersson, mcguire2019minimal, daramouskas2020methodology}; offloading computation to an external base-station~\cite{candan2018design, dunkley14iros, anwar2020autonomous}; or extending the onboard computing device either with general-purpose visual navigation engines~\cite{palossi2019DCOSS,palossi2021fully}
or with application-specific processors~\cite{navion2019,cnnslam2019,chen_slam2020,tinoosh_edge_ai,tinoosh_manycore}.
From the first category, Lambert~et~al.~\cite{lambert} exploit the STM32F4 MCU to implement a simple DL-based \textit{flight-controller} for hovering on a Crazyflie 2.0.
Similarly, Guanya Shi~et.~al.\cite{shi_neural_swarm} apply a simple DNN-based controller ($\sim\SI{27}{\kilo MAC}$) on a swarm of nano-drones, which enables multiple drones to fly safely in close proximity. % they use crazyflies 2.0
Zhao et al.~\cite{zhao2020learning} implement a CNN to improve the localization accuracy of a nano-UAV by modeling the sensor biases of the localization system.
This last DL model can run at \SI{200}{\hertz} on the onboard ARM Cortex M4 MCU requiring about \SI{2.7}{\kilo MAC/frame}, i.e., 10000$\times$ less operations than existing SoA CNN-based autonomous navigation workloads, e.g., $\sim$\SI{40}{\mega MAC/frame} in~\cite{palossi2019IOTJ}.
In~\cite{mcguire2019minimal}, the authors propose a lightweight navigation algorithm that enables a swarm of drones to explore an indoor area while avoiding collision with the walls by exploiting four laser distance sensors on each drone.
The autonomous navigation is commanded by a finite state machine, which maps the sensor output into drone control commands.
The simple sensor input (i.e., four single laser beams) results in an avoidance mechanism that can only detect large and homogeneous obstacles.
In contrast, our solution is based on the visual cues of the environment and, therefore, more general and robust against various types of obstacles.
\cite{daramouskas2020methodology} proposes an approach for navigation with obstacle avoidance using a random forest classifier.
They report a classification accuracy of $90\%$ while using a size of 229 nodes for the decision tree.
However, their approach was only developed and tested with synthetic data generated with the aid of a simulator.
While useful for some specific tasks, these tiny models are not a viable solution for more challenging navigation problems, like the ones we tackle in this work.

An approach to overcome the computational limitations of single-core MCUs is to offload intensive computation to off-board, wireless-connected computing resources. 
For example, in~\cite{anwar2020autonomous} the authors propose an autonomous navigation approach based on a CNN that uses reinforcement learning to adjust part of its parameters online.
The initial values of the weights are obtained by training the whole network with synthetic images obtained from a game engine, and they also prove the in-field functionality using a \SI{80}{\gram} drone.
However, the action space of their algorithm is limited to: move \SI{50}{\meter} forward, steer \SI{45}{\degree} and steer \SI{-45}{\degree}.
This results in less smooth and flexible navigation than our approach, which provides a continuous output for the steering angle and adaptive forward velocity.
In~\cite{candan2018design} the authors implement a fuzzy logic position controller and vision-based position estimation by offloading all the computation to an Intel i7 processor streaming images with a \SI{2.4}{\giga \hertz} radio.
This class of approaches, however, suffers from several important drawbacks~\cite{vulnerability_iot}: \textit{i}) it introduces network-dependent latency, which prevents the drone to operate farther than few tens of meters from the remote base-station, \textit{ii}) the noise on the transmission channel affects the reliability of the transmitted data, \textit{iii}) security becomes a concern for eavesdropping of confidential images and data and for denial-of-service attacks on the wireless connections, and \textit{iv}) the power consumption of the high-frequency radio transmission is significant and the wireless transceiver may dominate the power budget for control.

\subsubsection*{Accelerated nano-size UAVs}
A possible solution to the limitations imposed by single-core MCUs is
augmenting nano-UAVs with better compute functionality.
For larger UAVs, this is a common choice -- using devices such as NVIDIA GPUs, Intel Myriad, or Google Edge TPU\cite{Smolyanskiy,weedNet,movidius_myriad,google_tpu_vs_myriad} that are both flexible and highly efficient.
For nano-UAVs, however, the possibilities are more limited.
Some recent works emphasize the efficacy of application-specific integrated circuits (ASIC), which suitable for autonomous navigation functionalities~\cite{tinoosh_edge_ai,tinoosh_manycore,navion2019,cnnslam2019,chen_slam2020} on an low-power budget. 
Some of these systems have been designed to tackle specific UAV applications, such as visual-inertial odometry (VIO)~\cite{navion2019} and simultaneous localization-and-mapping (SLAM)~\cite{cnnslam2019,chen_slam2020}, within a power envelope of few hundred \SI{}{\milli \watt}. 
While extremely efficient, these systems are inflexible and they do not implement end-to-end flying functionality, but only accelerate some sub-functions, requiring anyways a mission and flight controller MCU.

The approach we follow in this work targets end-to-end mission control using a fully programmable parallel computing accelerator~\cite{conti_hetero}.
Parallel ultra-low-power (PULP) processors use small-scale multi-core clusters with 4-16 cores, with an enhanced RISC-V instruction set architecture (ISA), to exploit intrinsic parallelism of vision workloads, including CNNs.
An example of this new generation of energy-efficient devices is the 9-core GWT GAP8 SoC, which has been already applied to nano-drones~\cite{palossi2019IOTJ,palossi2019DCOSS,palossi2021fully} for tasks such as obstacle avoidance, lane detection, and pose estimation using CNNs in the range of 10-\SI{100}{\mega MAC/frame}. 
The key advantage of this approach is its flexibility and the capability to handle the end-to-end flight control task. 

% 4) Automatic deployment tools.
\subsubsection*{Automatic deployment tools}
Deploying multi-MMAC CNNs on an MCU-class device requires coping with a power envelope of less than \SI{1}{\watt}, a memory of just a few \SI{}{\mega\byte} or less, and limited peak performance, demanding for a strict co-optimization of the algorithmic, software, and hardware components~\cite{burrello2020dory,rusci2019memorydriven}.
Minimization of a DL model can be performed either with specific topological choices, like using depth-wise convolutions~\cite{mobilenetv2,mobilenetv3} or using quantization as a compression technique~\cite{choi2018pact,jacob_quantization} from \textit{float32} down to \textit{int8} or less, with a net 4$\times$ reduction of model footprint.
Quantization can also expose more data parallelism exploitable by packed-SIMD instructions~\cite{GautschinearthresholdRISCVcore2016}, improving the final inference throughput and the energy consumption.

Moreover, given a size-optimized network, the deployment challenge must be addressed, which consists in achieving the maximum utilization of computing resources by \textit{i}) parallelizing computation, \textit{ii}) managing the memory hierarchy (i.e., topology-dependent tiling), and \textit{iii}) minimizing data transfers overheads.
This is a key step especially for MCU devices, where the processing units are scarce by definition~\cite{burrello2020dory}.

General-purpose tools such as TFLite for MCUs and Larq~\cite{david2021tensorflow,larq,lai2018cmsisnn}, as well as vendor-locked tools like STM32 X-CUBE-AI\footnote{\url{https://www.st.com/en/embedded-software/x-cube-ai.html}} have been proposed to ease deployment on MCUs.
For PULP platforms, on which we focus in this work, two deployment tools have been recently introduced: GWT's \textit{AutoTiler}\footnote{\url{https://greenwaves-technologies.com/manuals/BUILD/AUTOTILER/html/index.html}}, which is partially closed-source, and DORY~\cite{burrello2020dory} with PULP-NN backend~\cite{pulp_nn}, an alternative open-source academic framework.

In this work, we exploit these recent advancements to bring the deployment of DL-based visual navigation on nano-drones from handcrafted and hand-tuned deployment~\cite{palossi2019IOTJ} to a new streamlined, automated methodology.
We leverage DNN deployment frameworks by integrating them in our flow and we achieve significantly improved performance and energy efficiency on autonomous navigation DL workloads, improving the nano-UAV in-field behavior and freeing resources for even more complex missions and tasks.

\section{Background}\label{sec:background}

\subsection{Robotic platform and Hardware}

\begin{figure}[t]
\centering
\includegraphics[width=1\linewidth]{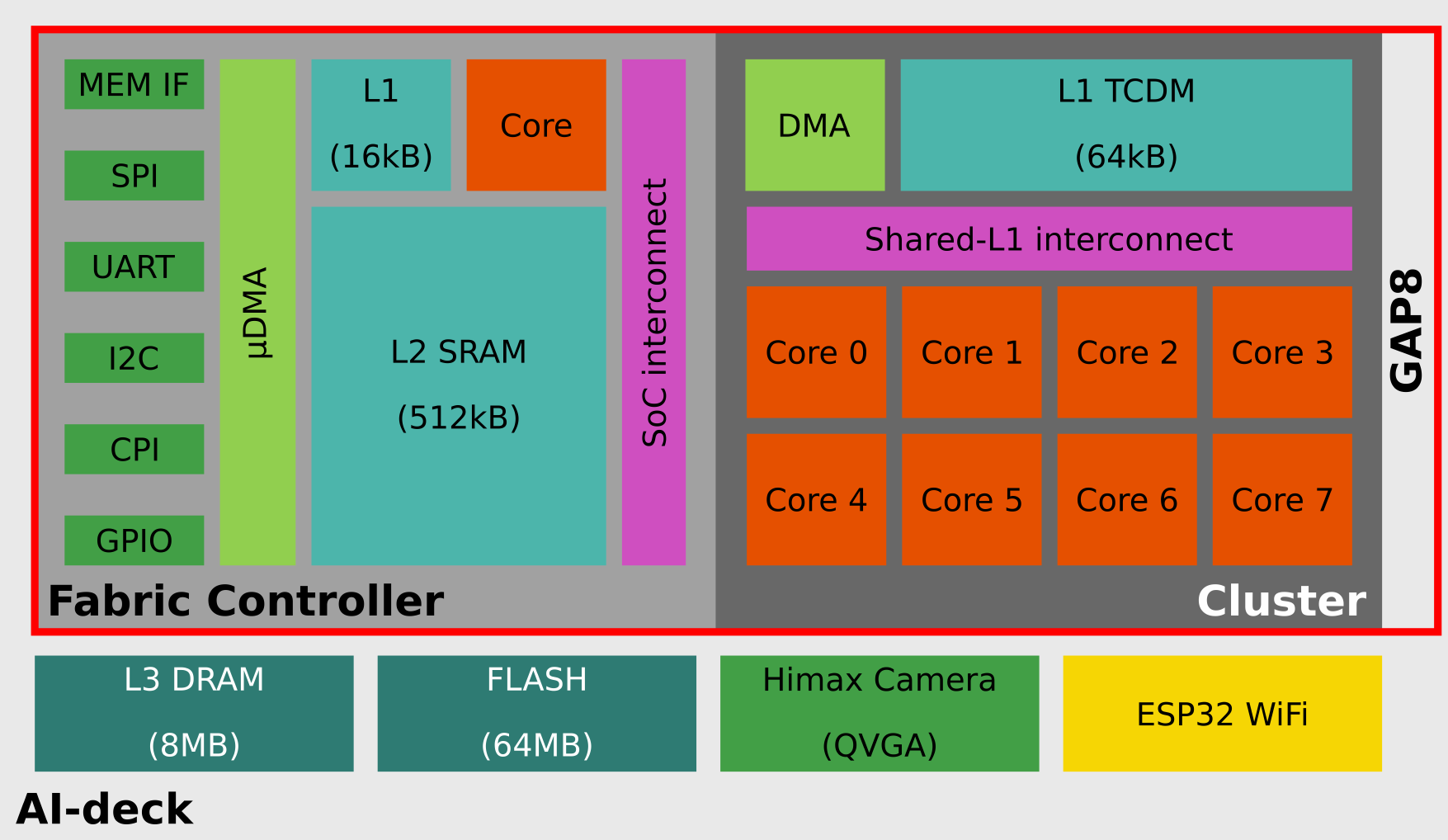}
\caption{AI-deck diagram and the GAP8 System-on-Chip architecture.}
\label{fig:gap8_archi}
\end{figure}

The algorithmic kernel of our application use case, i.e., PULP-Dronet V2, runs on a commercial embodiment of the PULP platform~\cite{GautschinearthresholdRISCVcore2016}, the GWT GAP8 SoC~\cite{gap8}.
GAP8 is a 1+8 general-purpose RISC-V-based multicore MCU, where the nine cores are organized in two power and frequency domains, namely the fabric controller (FC) and the cluster (CL), as shown in Figure~\ref{fig:gap8_archi}.
The former features one single core for control-oriented tasks, acting as an ``activity supervisor'' managing the interfaces to off-chip sensors/memories and orchestrating on-chip memory operations.
On the other hand, the CL is designed to execute computationally intensive parallel workloads, such as vision-based CNNs, so to enable high-level energy efficiency via the parallel computational paradigm~\cite{conti_hetero}.

All nine cores are based on the open-source RI5CY core~\cite{GautschinearthresholdRISCVcore2016}, which implements the \texttt{RV32IMC} instruction set and the \texttt{Xpulpv2} extension with digital signal processing (DSP) instructions: register-register multiply-accumulate, hardware loops, load/store operations with post-increment, packed single-instruction-multiple-data (SIMD), and specialized instructions for dot-product -- the latter two operating on vectors of 8-bit or 16-bit data.
The on-chip memory hierarchy is organized with \SI{512}{\kilo\byte} of L2 SRAM and \SI{16}{\kilo\byte} of L1 on the FC, while the CL has a \SI{64}{\kilo\byte} shared L1 as a tightly-coupled data memory (TCDM).
The TCDM is organized in sixteen banks, providing an aggregate bandwidth of \SI{11.2}{\giga\byte/\second}@\SI{175}{\mega\hertz}.
It is connected to the cores by means of a fully combinational logarithmic interconnect, which guarantees 0-wait state access from the cores in the absence of collisions -- in which case, one of the colliding accesses is stalled for one cycle.
To enable data transfer, the GAP8 SoC features two DMA engines: the first one, called $\mu DMA$, is in charge of data exchange with external I/O peripherals (e.g., DRAM, cameras, etc.) across a wide range of interfaces (e.g., QSPI, HyperBus, etc.).
Inside the CL domain, the second DMA controller is connected on one side to the TCDM logarithmic interconnected; on the other side to the AXI interconnect.
The CL DMA can be used to transfer data between the L2 and the L1 TCDM memory at up to 8B/cycle in either direction, guaranteeing data availability for the eight CL cores.

The robotic platform we employ in this work is the COTS open-source Crazyflie 2.1 nano-quadrotor from Bitcraze\footnote{https://www.bitcraze.io/products/crazyflie-2-1}.
This tiny UAV weighs \SI{27}{\gram}, has a diameter of $\sim$\SI{10}{\centi\metre}, and a total payload of $\sim$ \SI{15}{\gram}.
The underlying robotic platform is built around the STM32F405 MCU, in charge of all low-level flight controller functionalities, such as sensors' interfacing, state estimation, and low-level control. 

In this work, we employ a configuration that extends the basic nano-drone with two commercially available open-source, pluggable printed circuit boards (PCBs): the Flow-deck and the AI-deck~\footnote{https://store.bitcraze.io/products/ai-deck}.
The former features a low-resolution down-looking optical-flow camera coupled with a time-of-flight (ToF) sensor, enabling the drone to detect motions in any direction and providing a distance measurement from the ground, respectively.
The latter is the commercially supported version of the PULP-Shield research prototype, introduced in~\cite{palossi2019DCOSS}.
This board, shown in Figure~\ref{fig:gap8_archi}, extends the nano-drone's onboard capabilities with an energy-efficient GAP8 processor, off-chip DRAM, and Flash memory (\SI{8}{\mega\byte} and \SI{64}{\mega\byte}, respectively), a QVGA resolution low-power gray-scale camera (i.e., Himax HM01B0 sensor), and a versatile ESP32-based WiFi module\footnote{We remind the reader that, in this work, the Wi-Fi radio has been used only for debugging and showcasing video-streaming purposes.}.

Combining the STM32 MCU with the GAP8 SoC enables the heterogeneous architectural paradigm at the ultra-low-power scale~\cite{conti_hetero}, enabling onboard execution of sophisticated vision-based algorithms.
In this \textit{host-accelerator} context, the STM32 represents the \textit{host}, handling control-oriented tasks (i.e., flight controller), while the GAP8 offers general-purpose parallel computation capabilities, acting as the \textit{accelerator} for compute-intensive perception and navigation tasks.

\subsection{PULP-Dronet CNN}

Dronet~\cite{loquercio2018dronet} is a vision-based end-to-end autonomous drone navigation CNN, deployed on a nano-drone, for the first time, in the seminal PULP-Dronet project~\cite{palossi2019IOTJ}.
This shallow NN is based on three consecutive ResNet~\cite{resnet2015} blocks that branch the last layer to produce two outputs: a probability of collision (classification problem) and a steering angle (regression problem). 
The CNN was originally developed using 16-bit fixed-point arithmetic as the result of a quantization-aware training process.
Images used in the original training/validation/testing, and also in this work, are partitioned in three disjoint sets: 
\begin{itemize}
 \item \textbf{Udacity}: $\sim$39.1K high-resolution images labeled only with steering angle.
 \item \textbf{Bicycle}: $\sim$32.2K high-resolution images labeled only with collision probabilities. 
 \item \textbf{Himax}: $\sim$1.3K low-resolution images collected from the same camera aboard our target nano-drone and labeled only with collision probabilities.
\end{itemize}
The union of Udacity and Bicycle sets results in the so-called \textit{Original} dataset that we use to train our PULP-DroNet V2 in PyTorch (100 epochs) and to select the models that minimize both regression and classification error on the validation set.

\section{Deployment Automation Flow}\label{sec:methodologies}

The development of AI-based algorithms on MCU-class processors, aboard a nano-drone, is a complex multi-objective optimization problem that must take into account: \textit{i}) memory availability, \textit{ii}) power envelope, \textit{iii}) hardware limitations (e.g., no FPU), and \textit{iv}) throughput.
Therefore, to enable the execution of PULP-Dronet on GAP8 under these constraints, we assemble and streamline a flow of automated tools that divide the process into two main stages: \textit{i}) quantization of the neural network,
and \textit{ii}) hardware-aware deployment of the quantized model.

\subsection{Quantization}
This stage, remapping the CNN's numerical representation, e.g., from \texttt{float32} to \texttt{int8}, enables efficient integer computation on the underlying hardware.
From a mathematical viewpoint, the tools we consider in this work focus on \textit{uniform affine quantization}: all tensors $\mathbf{t}$ (typically inputs $\mathbf{x}$, outputs $\mathbf{y}$ or weights $\mathbf{w}$) are first restricted to a known range $[\alpha_t, \beta_t)$, then they are mapped to $N$-bit purely integer tensors $\widehat{\mathbf{t}}$ by means of a bijection:
\begin{equation}
 \mathbf{t} = \alpha_\mathbf{t} + \varepsilon_\mathbf{t}\cdot \widehat{\mathbf{t}} \label{eq:1}\;, 
\end{equation}
where $\varepsilon_\mathbf{t} = (\beta_\mathbf{t}-\alpha_\mathbf{t}) / (2^{N}-1)$.
$\varepsilon_\mathbf{t}$ is often called the \textit{scaling factor} used to scale tensors from their floating-point to their integer representation. 
Quantization flows enforce the representation quantized tensors of all waits and part of the data tensors in the network -- the latter typically together with ReLU activation functions.

NNTOOL is the NN mapping flow developed by GWT, included in the GAP\textit{flow}, that converts a TFLite topology graph into a new custom representation.
It is distributed as part of the GAP8 software development kit\footnote{\url{https://github.com/GreenWaves-Technologies/gap_sdk}}.
NNTOOL performs ``layer-fusion'', post-training calibration and quantization (8/16-bit), and folds batch normalization (BN) into the convolution layer that precedes it, avoiding costly intermediate buffers and saving a small amount of memory traffic (i.e., \SI{1.792}{\kilo\byte} in Dronet).
On the other hand, NEMO is the quantization tool used by the open-source pipeline NEMO/DORY~\cite{burrello2020dory}, which provides both post-training quantization (i.e., quantizing the model without further re-training, using only lightweight calibration) and quantization-aware training (i.e., quantization at training-time, to mitigate potential accuracy loss).
NEMO does not fold BN layers, but instead, it converts them into fully integer channel-wise scaling operations~\cite{rusci2019memorydriven}. 

\begin{figure}[t]
\centering
\includegraphics[width=1\linewidth]{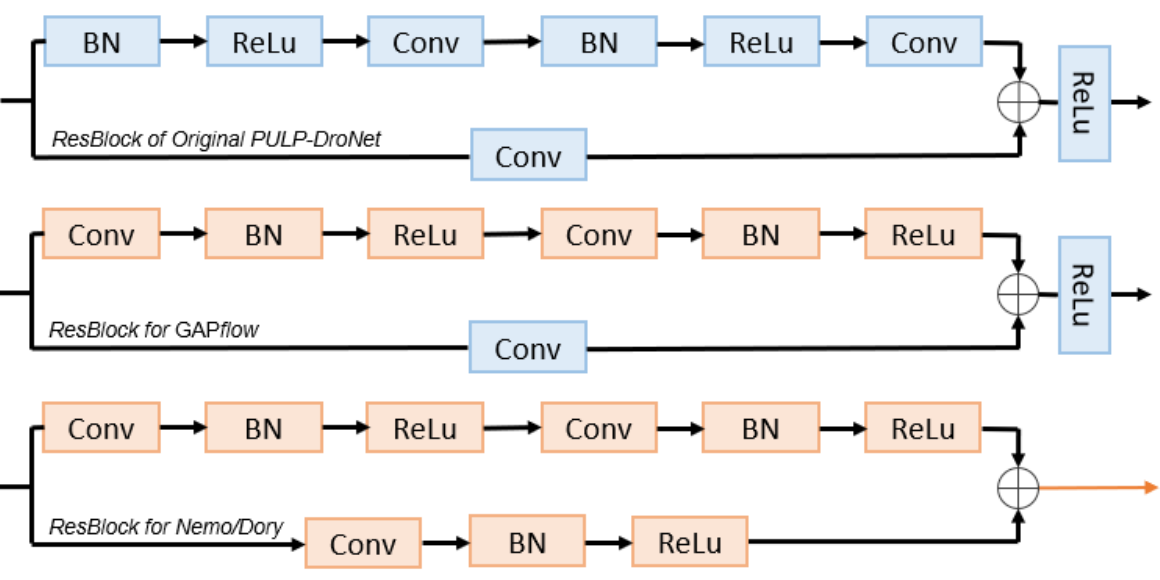}
\caption{ResBlocks of PULP-DronetV1, PULP-DronetV2 GAP\textit{flow} and PULP-DronetV2 NEMO/DORY (top-bottom order).}
\label{fig:resblocks}
\end{figure}

For our application, we apply post-training quantization at 8-bit for both NEMO and NNTOOL, which is -- to date -- the most commonly adopted quantized bit-width and is supported by both flows.
Specifically, NNTOOL employs a signed \textit{int8} format for both activations and weights, whereas NEMO employs \textit{uint8} for activations and \textit{int8} for weights.
Both tool-sets require a \textit{Conv-BN-ReLU} pattern for all main branches of each ResBlock.
This simplifies both quantization and deployment: the accumulated tensor at the output of the Conv operation naturally requires a finer grain representation than that of inputs and weights -- both flows employ 32 bits.
Integer scaling and ReLU can be applied to a single element at a time, meaning that there is no need to materialize a full tensor of 32 bits elements -- rather, each element is produced at 32 bits by Conv but immediately reduced to 8 bits by ReLU or BN+ReLU.
The baseline version of the NEMO flow does not support the quantization of data that is not at the output of a ReLU; as a consequence, we introduce a further modification by pushing the final ReLU of the ResBlocks back to the residual branch. 
Figure~\ref{fig:resblocks} summarizes the minor modifications that were used in the two flows with respect to PULP-DronetV1 -- establishing two new NN topologies, namely, PULP-DronetV2 GAP\textit{flow} and PULP-DronetV2 NEMO/DORY.

\subsection{Hardware-aware deployment} \label{sec:closed-loop}
The deployment goal is to enable and exploit the hardware platform by generating C code that: \textit{i}) maximizes the parallel execution over all available cores, and \textit{ii}) minimizes the data transfers overhead.
On GAP8, the main challenge is the limited L1 memory (\SI{64}{\kilo \byte}), that forces the deployment tools to solve an optimization problem, partitioning the tensors into smaller chunks of data, called tiles, to be moved between L2 and the L1 memory.

Both GAP\textit{flow} and NEMO/DORY partition this problem in two separate parts: \textit{i)} a set of optimized kernels operating exclusively on L1 data tiles, and \textit{ii)} a tiling solver to define the optimal size for tiles and generate the code for the related data transfers between L2 and L1, including double buffering for all tensors.
As optimized primitives, GAP\textit{flow} relies on a set of open-source kernels available within the GAP SDK, with the possibility of defining custom ones.
NEMO/DORY uses the open-source PULP-NN library\footnote{\url{https://github.com/pulp-platform/pulp-nn}}~\cite{pulp_nn}.
The tiling solver employed by the GAP\textit{flow} is a proprietary tool called AutoTiler, whereas NEMO/DORY employs  DORY, an open-source flow~\cite{burrello2020dory}.

% HWC vs CHW
The two primitive libraries exploit different data layouts, affecting the final performance on the Conv layers.
PULP-NN, employed by NEMO/DORY, exploits the height-width-channel (HWC) layout, where the data along the channels' dimension is stored with a stride of one, while the data along the width dimension is stored with a stride equal to the number of channels.
NNTOOL uses the channel-height-width (CHW) format, reverting the previous order.
The convolutional layer can be performed either as a \textit{direct convolution} or as a \textit{matrix-matrix multiplication}, optimized for CHW and HWC layouts, respectively.
Both implementations have pros and cons.
Direct convolution uses a sliding window with a masking/shuffling mechanism, while in the matrix multiplication case, we need to pay some extra overhead to rearrange the input data to a single-dimension tensor (i.e., \texttt{im2col}, image-to-column) so that the convolution can be computed as matrix multiplication.
On the other hand, matrix multiplication is a more regular operation than convolution, it is essentially identical for any filter size, and it does not require any data shuffling.
In general, the HWC layout and the matrix-matrix multiplication become more convenient when the feature map of the convolved layer has many input channels.
Conversely, the CHW data layout used by the GAP\textit{flow} is most advantageous with direct convolutions on Conv layers with spatial dimensions much larger than the number of input channels.

The tiling solver employed by GAP\textit{flow} is the AutoTiler, whereas the open-source flow employes DORY~\cite{burrello2020dory}.
AutoTiler is a proprietary partially closed-source tool.
It can automatically promote full tensors from L3 to L2 and to L1 or tile them in order to maximize performance using the GAP\textit{flow} backend primitives.
On the other hand, DORY specifies tiling as two separate problems -- one for L3/L2, the other for L2/L1 transfers.
To promote data from L3 to L2, DORY uses a set of simple heuristics, such as looking at the known-good solution first (e.g., copying the full weights for the next layer in L2 while the current one is being run; keep all activations in L2) and revert to less optimal ones when the former ones are not feasible (e.g., move part of the activations in L3).
For the L2/L1 transfers, insisting on a much smaller L1 size (64 KB), tiling is specified as a constrained optimization problem with the objective to maximize L1 utilization, and at the same time maximize a few hardware-aware heuristics (e.g., favor tiles that are better parallelized due to their specific sizes).

Overall, we observe that for our CNN the AutoTiler finds a better solution for layers that are spatially large and without many input channels, such as the first convolutional layer; DORY, on the other hand, performs better for layers where the number of input channels is high.
By inspection of generated code, we also notice that the AutoTiler is able to fuse consecutive layers (e.g., convolutions, max-pooling, and ReLU) and apply multiple operations directly on the same L1 tile, avoiding an intermediate copy to L2.
This is the case for the first part of PULP-Dronet (i.e., Conv$5\times5$ + MaxPool), where the AutoTiler merges the first two layers, while DORY executes them one after the other, as it can not store in the L1 memory all the needed parameters required by the HWC layout.
Both GAP\textit{flow} and NEMO/DORY implement, whenever is possible, pipelined memory/computation phases by means of the GAP8's $\mu DMA$ (L3-L2).

\subsection{Platform integration \& low-level control}

\begin{figure}[t]
\centering
\includegraphics[width=1\linewidth]{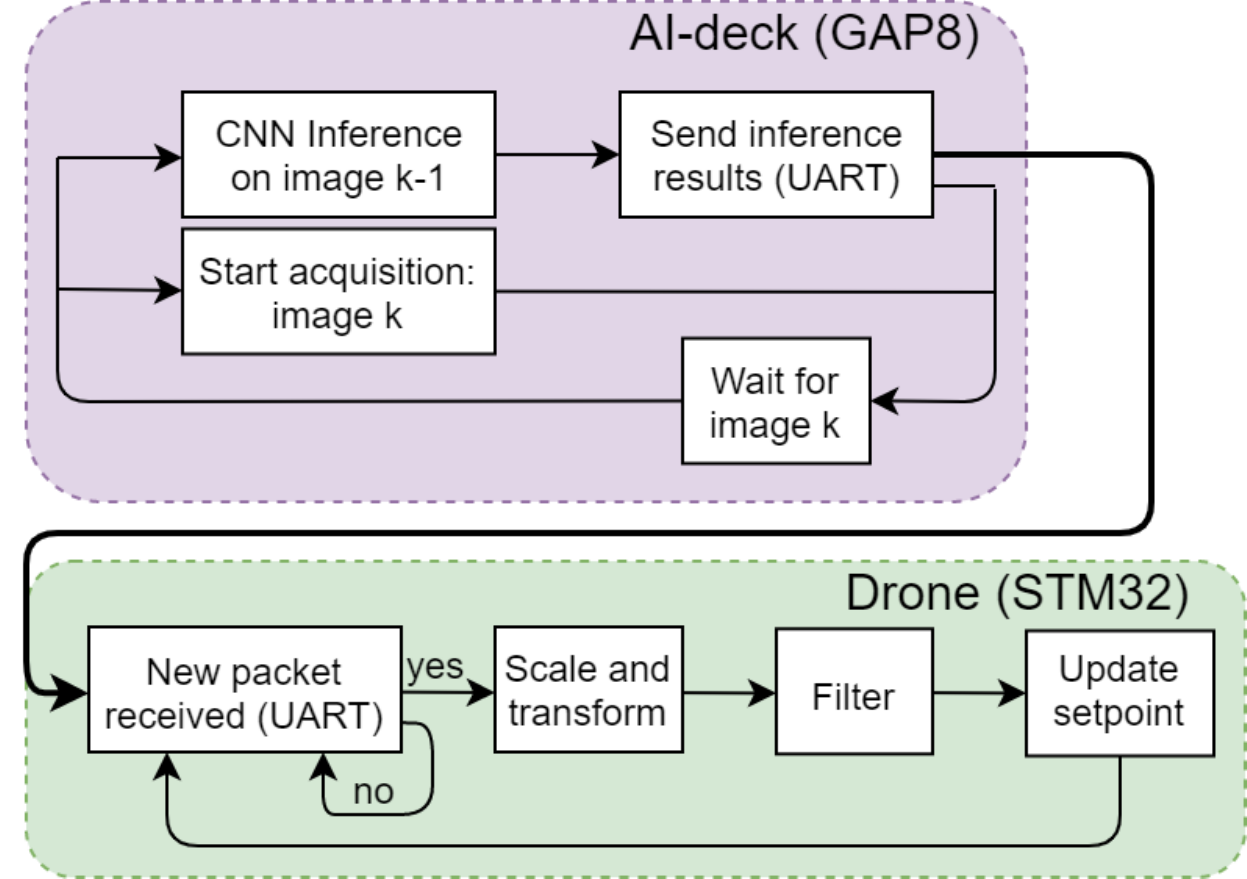}
\caption{Overview of the main acquisition and control loops. The AI-deck is in charge of image acquisition and perception, and the drone's MCU runs the application that interprets the perception results and transforms them into flight commands}.
\label{fig:per_ctrl}
\end{figure}

% AI-deck
To enable autonomous navigation on the nano-drone, the inference results of the CNN running in the AI-deck have to be communicated to the flight controller, running on the Crazyflie's main board.
This controller is in charge of running control algorithms that drive the drone and run on top of the STM32F405 MCU.
Communication between the Crazyflie flight controller and the AI-deck happens via UART communication at 115200 baud.
Figure~\ref{fig:per_ctrl} shows the stages of the perception and control.
In the AI-deck side (purple), whenever a new inference is started for the current image (k) in the CL, the FC also commands the acquisition of the next image (k+1), which will serve as input for the next inference. 
Every time a new inference result is available, the AI-deck sends this data via UART.
When using GAP\textit{flow}, the outputs of the deployed CNN are also quantized at 8 bits: the inference result, therefore, simply consists of 2 bytes - one for the probability of collision and one for the steering rate.
When using NEMO/DORY, the accumulated values after the final layer -- represented as 32 bit integers -- are directly used as outputs.
In this case, the inference result is sent as a packet of 8 bytes.

\begin{lstlisting}[
  style=mystyle,
  float=tb,
  caption={The listing describes the data processing that is applied to the raw output of the CNN to obtain the setpoint that is communicated to the drone's commander.},
  label={lst:processing_flow}
]

while 1:
  if uart_data_available:
    # reset the flag
    |uart\_data\_available $\leftarrow$ 0|
    # scale data
    |$p\sb{col} \leftarrow data[0] * c\sb{scale0}$|
    |$\omega\sb{steer} \leftarrow data[1] * c\sb{scale1}$|
    # compute the integral
    |$I(k) \leftarrow I(k+1) + (p\sb{col}-0.3)$|
    |$I(k) \leftarrow \mathrm{clip}_{[0,3)}(I(k))$|
    # transform: compute forward velocity
    |$p\sb{col} \leftarrow p\sb{col} + w \cdot I(k)$|
    |$v\sb{unfilt} \leftarrow v\sb{target}(k)\cdot(1-p\sb{col})\sp{2}$|
    # filter forward velocity
    |$v\sb{set}(k) \leftarrow \alpha\sb{1} \cdot v\sb{unfilt} + (1 - \alpha\sb{1}) \cdot v\sb{set}(k-1)$|
    # filter steering rate
    |$\omega\sb{set}(k) \leftarrow \alpha\sb{2} \cdot \omega\sb{steer} + (1 - \alpha\sb{2}) \cdot \omega\sb{set}(k-1)$|
    # command the drone
    |command($v\sb{set}(k)$, $\omega\sb{set}(k)$)|
\end{lstlisting}

% STM32
The program that allows receiving the data from the AI-deck and processing it to drive the drone is integrated as a new task in the drone's firmware. 
To achieve a computational-efficient data exchange, the drone's MCU uses a DMA mechanism to receive the UART data from the AI-deck.
The DMA is configured to trigger an interrupt whenever a certain number of bytes has been received -- which is two in our case.
When the interrupt is triggered, a binary flag (i.e., 0 or 1) that indicates new data available from UART is set.
The main loop of the application evaluates the value of this flag every \SI{5}{\milli \second}, and in case it is set, it reads the two received bytes and then resets the flag back to 0.

The main overview of the inference post-processing stages is given in Figure~\ref{fig:per_ctrl}, and each step (scale, transform, filter) is detailed by Listing~\ref{lst:processing_flow}.
First, the two pieces of data ($data[0]$ and $data[1]$) associated with the output of the CNN are dequantized by multiplying them by the scaling constants resulting from the quantization process.
The scaling constants are programmed in the drone's MCU firmware.
Next, $I(k)$ is computed, which is an integral term that is added to the probability of collision ($p\sb{col}$), and it is meant to penalize the lasting effect of the obstacles in the field view.
We noticed that the CNN is sometimes unsure about particular frontal obstacles, and the probability of collision oscillates from values $>0.8$ to values below $0.3$.
Thanks to the addition of the integral term, when the CNN is indicating an obstacle with a $p\sb{col}>0.3$, $I$ will increase over time, building up the drone's confidence that it is facing an actual obstacle. 
When the CNN's inference indicates an obstacle-free horizon ($p\sb{col}<0.3$) for a longer time again, $p\sb{col}-0.3$ is negative and therefore the integral decreases.
We clip the value of $I(k)$ to the interval $[0,3)$ as negative confidence is meaningless (on the lower side), and excessive confidence could result in a windup effect.
We scale $I(k)$ by a scaling constant $w$ that establishes how much impact the integral has on the final value of the probability of collision.
In our experiments, we set this value to $0.2$.

To convert the probability of collision $p\sb{col}$ into forward velocity ($v\sb{unfilt}$), we use a simple square low -- penalizing velocity quadratically with respect to $p\sb{col}$.
Furthermore, to reduce the high-frequency noise associated to $v\sb{unfilt}$, this value is filtered using a first-order, low-pass infinite impulse response (IIR) filter defined by the coefficient $\alpha\sb{1}$ (we use $\alpha\sb{1}=0.6$.
The same type of filter (defined by $\alpha\sb{2}$) is also used for the steering rate $\omega\sb{steer}$.
We use $\alpha\sb{2}=0.7$; we observed experimentally that a lower value results in an increased delay and in low-frequency oscillations around the navigation path.
The filtered values for the forward velocity and the steering rate ($v\sb{set}$ and $\omega\sb{set}$) represent the new flight setpoint, which is transmitted to the drone's flight controller. 

\section{Results}\label{sec:results}

In this section, we present three main classes of results: \textit{i}) regression and classification capability of the proposed PULP-Dronet V2 CNNs; \textit{ii}) onboard power analysis and inference performance; \textit{iii}) in-field closed-loop control accuracy and the real-time performance.

\begin{table}[tb]
\centering
\setlength{\tabcolsep}{0.7em}
\renewcommand{\arraystretch}{1.1}
\caption{Regression \& classification -- in \textbf{bold} our best fixed8 scores}
\label{tab:accuracy-regression-results}
\begin{tabular}{|c|c|cccc|} 
\hline
\multicolumn{2}{|c|}{\textbf{Training}} & \multicolumn{4}{c|}{\textbf{Testing}} \\ 
\hline
\multirow{2}{*}{\begin{tabular}[c]{@{}c@{}}NN\\topology\end{tabular}} & \multirow{2}{*}{Dataset} & \multirow{2}{*}{Precision} & \multicolumn{2}{c}{\textit{Original Dataset}} & \textit{Himax Dataset}\\ 
\cline{4-5}
 & & & RMSE & Acc & Acc \\ 
\hline
\multirow{4}{*}{V1} & \multirow{2}{*}{Original} & Float32 & 0.105 & 0.945 & 0.845 \\
 & & Fixed16 & 0.097 & 0.935 & 0.873 \\
\cline{2-6}
 & \multirow{2}{*}{\begin{tabular}[c]{@{}c@{}}Original\\ +Himax\end{tabular}} & Float32 & 0.109 & 0.964 & 0.900 \\
 & & Fixed16 & 0.110 & 0.977 & 0.891 \\
\hline
\multirow{4}{*}{\begin{tabular}[c]{@{}c@{}}V2\\ GAP\textit{flow}\end{tabular}} & \multirow{2}{*}{Original} & Float32 & 0.126 & 0.915 & 0.831 \\
 & & Fixed8 & \textbf{0.124 } & 0.916 & 0.840 \\
\cline{2-6}
 & \multirow{2}{*}{\begin{tabular}[c]{@{}c@{}}Original\\ +Himax\end{tabular}} & Float32 & 0.136 & 0.925 & 0.881 \\
 & & Fixed8 & 0.135 & \textbf{0.925 } & \textbf{0.886 } \\
\hline
\multirow{4}{*}{\begin{tabular}[c]{@{}c@{}}V2\\ NEMO/ \\DORY\end{tabular}} & \multirow{2}{*}{Original} & Float32 & 0.146 & 0.902 & 0.841 \\
 & & Fixed8 & 0.143 & \textbf{0.903 } & 0.836 \\
\cline{2-6}
 & \multirow{2}{*}{\begin{tabular}[c]{@{}c@{}}Original\\ +Himax\end{tabular}} & Float32 & 0.118 & 0.893 & 0.905 \\
 & & Fixed8 & \textbf{0.120} & 0.892 & \textbf{0.900} \\
\hline
\end{tabular}
\end{table}

\subsection{Regression \& classification performance}\label{sec:regression_classification}
In Table~\ref{tab:accuracy-regression-results}, the NNs quality metrics are reported as accuracy for the classification problem and root-mean-squared error (RMSE) for the regression one, using the Original and Himax datasets for both training and testing.
We also evaluate the impact of quantization w.r.t. floating-point calculation for each model proposed and each training set.
Particular attention should be given to the scores achieved on the Himax testing set as it maps the type of images available on our flying drone.

\subsubsection{Testing on Original Dataset}
When training on the Original dataset with a \texttt{float32} format, both the proposed models show a lower performance w.r.t. the PULP-Dronet V1.
The drop is between 0.02 and 0.04 in RMSE and up to $\sim 4\%$ of accuracy.
This small difference can be ascribed to \textit{i}) the differences in the NNs topologies and quantization factors (8-bits vs. 16-bits), and \textit{ii}) a weak CNN's convergence.
We attribute this weak convergence to the disjoint training datasets for the two problems (i.e., classification and regression); the relatively large unified CNN front-end, resulting in shared weights for two very different tasks up to the very last layer, may also contribute.
Our models reveal a different behavior when training on the Original+Himax dataset (\texttt{float32}) than the equivalent models trained on the Original set.
The GAP\textit{flow} model (similarly to the V1 baseline) slightly improves the classification performance ($\sim +1\%$ accuracy) at the price of a small reduction in the regression capability ($\sim +0.01$ RMSE); instead, the NEMO/DORY model shows the opposite behavior, i.e., accuracy $\sim -1\%$ and RMSE $\sim -0.02$.
The small differences are mainly because the two pipelines use slightly different topologies (Figure~\ref{fig:resblocks}), due to the different approach to batch normalization in the quantized regime. 

Our four 8-bit quantized models show a minimal variation in both RMSE and accuracy metrics (within 0.003 and $0.1\%$, respectively) compared to the respective \texttt{float32} version, which is not the case for the V1 baseline as it improves the RMSE of 0.008 and drops $1\%$ of accuracy.
The lower variance is the consequence of the different quantization schemes adopted.
In fact, in contrast to the quantization-aware training used in V1, we do not need to retrain the NNs to change the numerical domain due to our post-training quantization.
Moreover, post-training quantization does not require the model's training dataset, enabling a faster and straightforward process for producing the quantized model.

\subsubsection{Testing on Himax Dataset}
When training on the Original dataset with a \texttt{float32} format, all the three NN topologies score a similar accuracy of $\sim 83-84\%$.
Instead, training on the Original+Himax dataset, the accuracy of all models increases up to $\sim 88-90\%$, proving the beneficial effect of the dataset extension.
Finally, the 8-bit quantization of the V2 models does not affect the accuracy, keeping the same maximum ($90\%$) achieved by the 16-bit quantized baseline (V1), which again shows higher variance.
Ultimately, the proposed PULP-Dronet V2 models achieve the same accuracy of the V1 baseline on the in-field-collected Himax dataset, despite the reduced data-type (8-bit vs. 16-bit), leading to a highly desirable $2\times$ reduction in the memory footprint.
Including the Himax dataset in the training process does not aim at mitigating any quantization effect, but to ensure a better tuning between the CNN model and the onboard camera.
In conclusion, regardless of the testing dataset, the 8-bit quantization preserves the CNN's accuracy unaltered w.r.t to the \texttt{float32} representation and allows us to deploy and run our model on the target platform successfully.
\footnote{The model's memory footprint could be further reduced with stronger quantization, e.g., 4/2/1-bit; however, this approach is not guaranteed to be sufficient to maintain the full precision regression/classification performance as shown by our work and by the SoA when adopting 8-bit quantization~\cite{jacob2018quantization, choi2019accurate}.}

\subsection{Onboard performance}\label{sec:power}

\subsubsection{Power consumption \& inference performance}
\label{sec:CL_FC_power_consumption}
We evaluate the execution time and power traces of the proposed models running them on the GAP8 and using a RocketLogger data logger~\cite{sigrist2016rocketlogger} (\SI{64}{\kilo sp \second}).
For these experiments, the SoC's operating points are FC@\SI{50}{\mega\hertz}, CL@\SI{100}{\mega\hertz}, VDD@\SI{1}{\volt}, as the most energy-efficient configuration~\cite{palossi2019IOTJ}, and FC@\SI{250}{\mega\hertz}, CL@\SI{175}{\mega\hertz}, VDD@\SI{1.2}{\volt} to push the system at its maximum performance.
The GAP\textit{flow} model processes one frame in \SI{1.05}{\mega cycle}, while the NEMO/DORY model needs \SI{11}{\percent} fewer cycles.
Since our models use almost the same topology of PULP-Dronet V1, they all compute $\sim\SI{41}{\mega MAC/frame}$, as the original baseline.
We mention, however, that each MAC in PULP-Dronet V2 is an 8x8-bit MAC rather than a 16x16-bit MAC as in V1.

\begin{figure}[t]
\centering
\includegraphics[width=1\linewidth]{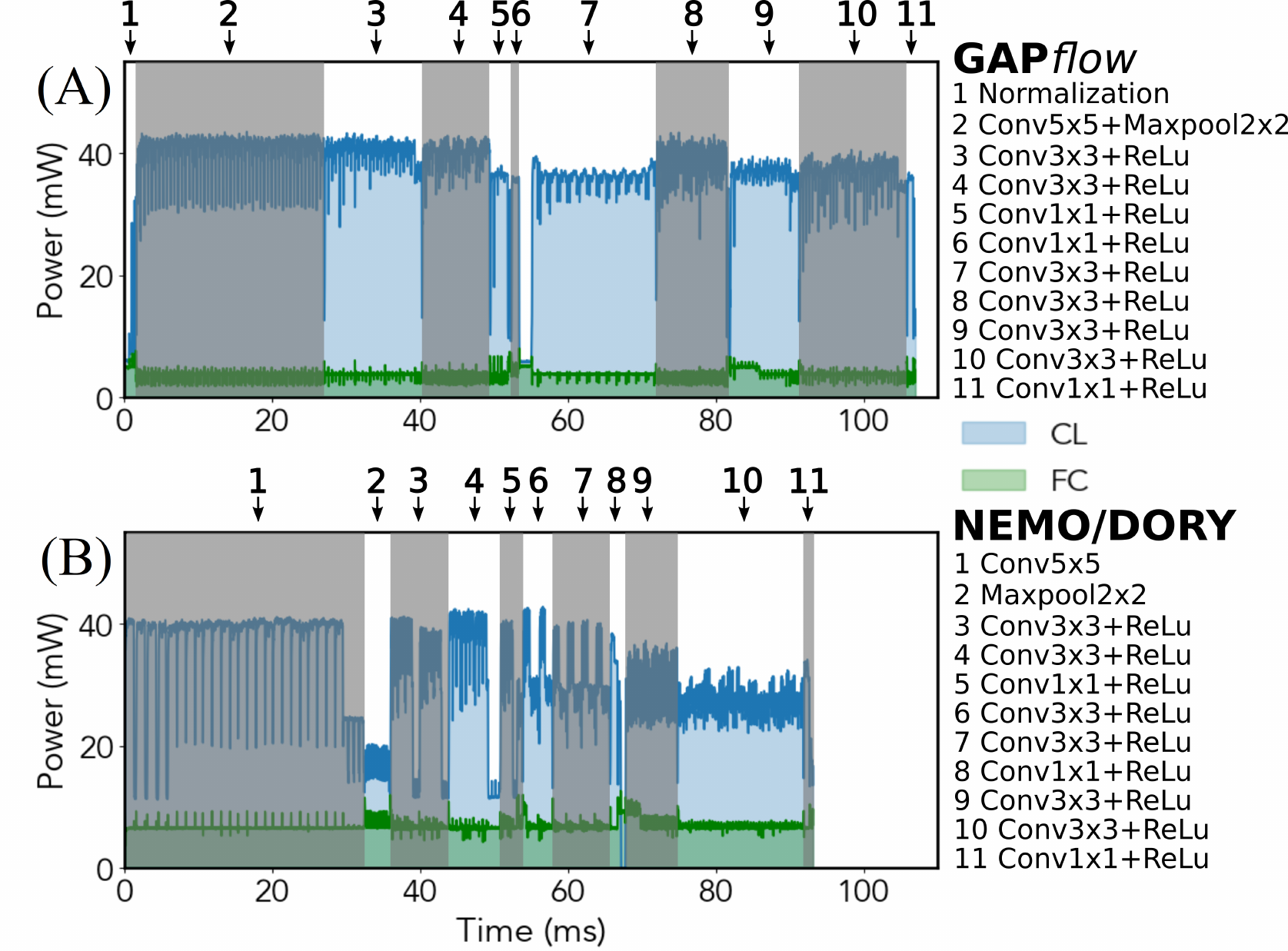}
\caption{GAP8's power waveforms for: FC@\SI{50}{\mega\hertz}, CL@\SI{100}{\mega\hertz}, VDD@\SI{1}{\volt}, i.e., the most energy efficient configuration.}
\label{fig:power_curves}
\end{figure}

Figure~\ref{fig:power_curves} shows the power consumption for one frame inference for both models (most energy-efficient configuration) and highlights the time intervals associated with the execution of each CNN's layer.
The GAP\textit{flow} model (Figure~\ref{fig:power_curves}-A) shows an initial extra stage to normalize the 8-bit input data-range to [-127,+128], as well as some cluster idleness at the begin of layer 7, due to a $\mu DMA$ transfer wait.
A major difference between the two models is visible in the first two layers (i.e., conv5$\times$5 and max-pool), as the GAP\textit{flow} model achieves better performance merging them (see Section~\ref{sec:methodologies}).
Nevertheless, NEMO/DORY outperforms its counterpart during the Conv+ReLU pattern, present in each ResNet block.

\begin{figure*}[t]
 \includegraphics[width=\textwidth]{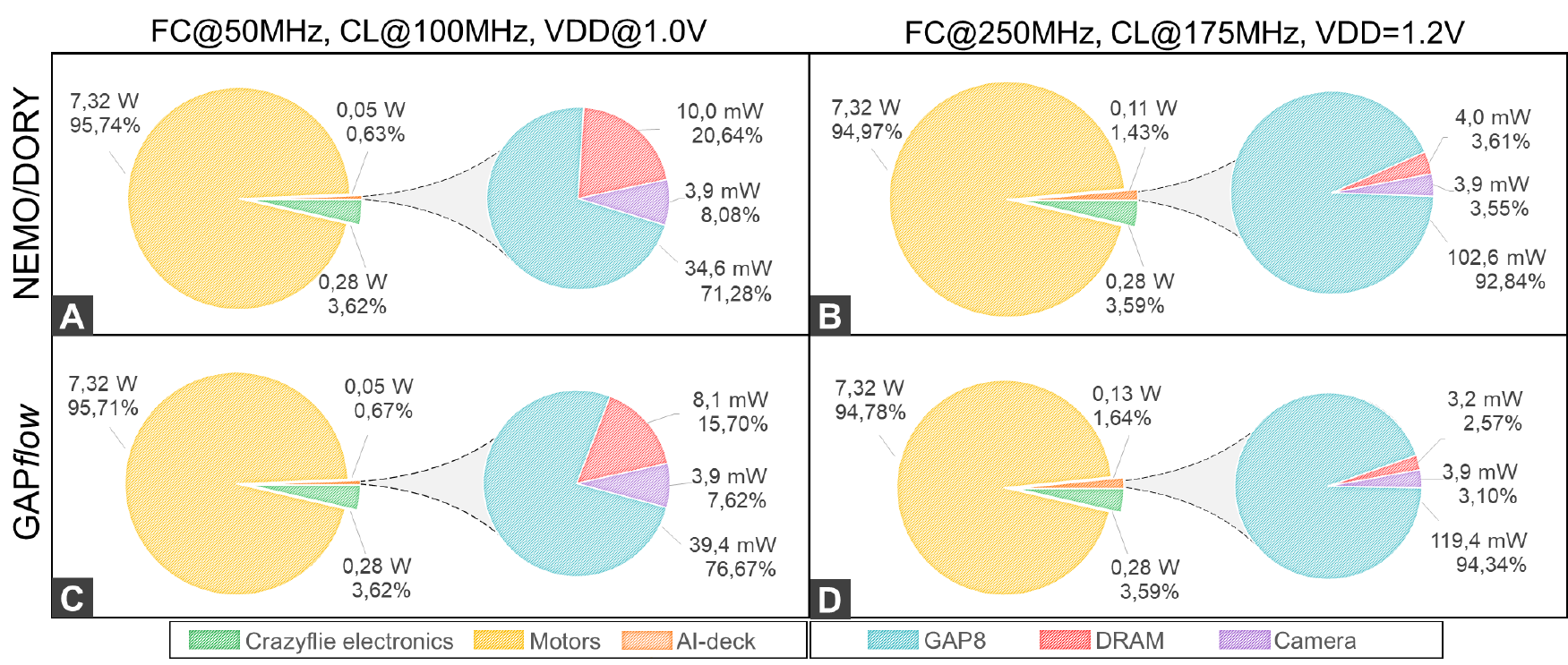}
 \caption{The nano-drone's power envelope break-down, with AI-deck zoom-in. A/B) NEMO/DORY, and C/D) GAP\textit{flow} framework. SoC running at FC@\SI{50}{\mega\hertz}, CL@\SI{100}{\mega\hertz} (A/C) and FC@\SI{250}{\mega\hertz}, CL@\SI{175}{\mega\hertz} (B/D), the most energy efficient and maximum performance configurations, respectively.}
 \label{fig:power_breakdown}
\end{figure*}

In the most energy-efficient configuration, the GAP\textit{flow} and NEMO/DORY models achieve similar performance for one frame inference, as $\sim\SI{9}{frame/\second}$ @ $\sim\SI{40}{\milli\watt}$ and $\sim\SI{10}{frame/\second}$ @ $\sim\SI{35}{\milli\watt}$, respectively, improving the throughput vs. PULP-Dronet V1 ($40\% - 60\%$).
Running the same test, with the SoC's maximum frequencies, the GAP\textit{flow} model scores $\sim\SI{17}{frame/\second}$ @ $\sim\SI{119}{\milli\watt}$, while the NEMO/DORY one peaks at $\sim\SI{19}{frame/\second}$ @ $\sim\SI{102}{\milli\watt}$.
Even if the GAP\textit{flow} model exposes a more balanced utilization of the available cores, i.e., almost constant CL power consumption, it pays the overhead for the CWH layout applied to small WH.
Conversely, DORY, even with a less balanced parallel workload, i.e., scattered profile inside layers 3, 4, 5, 6, and 7, reduces overheads due to its HWC layout.

Ultimately, quantization is a key-enabler technique to fully deploy a DNN model on resource-constrained COTS MCUs, which usually lack floating-point units (e.g., the GAP8 SoC). 
For the same reason, it is hard to precisely compare the execution performances of a full-precision model vs. a quantized one on such a processor.
An option is represented by soft-float emulation of all floating-point operations; although, this approach would introduce a major execution overhead.
Therefore, we show how quantization improves memory footprint and inference throughput by comparing the proposed 8-bit model to the quantized V1 baseline (16-bit).
On the one hand, the 2$\times$ reduction in the data-type format halves the total size of parameters from \SI{0.64}{\mega\byte} to \SI{0.32}{\mega\byte}. 
On the other hand, it allows for efficient exploitation of the GAP8's SIMD instructions, resulting in a throughput speedup of 1.5-1.6$\times$ w.r.t. the baseline. 
This mismatch in speedups (i.e., memory footprint and throughput gain) can be ascribed to multiple factors, such as \textit{i)} non-MAC and non-accelerable operations, and \textit{ii)} non-idealities, e.g., imbalanced workload and marshaling overheads.

\subsubsection{State-of-the-Art comparison}
We validate the two proposed GAP8's pipelines, comparing their performance against one of the most popular CNN libraries for MCUs: CMSIS-NN~\cite{lai2018cmsisnn}.
CMSIS-NN peaks at \SI{0.71}{MAC/cycle} on 8-bit data convolutions~\cite{lai2018cmsisnn} on a CNN's layer similar to our $3\times3$ convolution, for which we achieve at best \SI{0.81}{MAC/cycle/core} and \SI{1.0}{MAC/cycle/core} for the GAP\textit{flow} and NEMO/DORY, respectively.
Considering all the inevitable non-idealities, such as sub-optimal load balancing, we yield a weighted performance, for the entire CNN, of \SI{3.9}{MAC/cycle} (GAP\textit{flow}) and \SI{4.7}{MAC/cycle} (NEMO/DORY), employing all the GAP8's cores.
To concertize this comparison, we consider the CMSIS-NN on top of a high-performance Cortex-M7-based single-core STM32H723VE, which can achieve up to \SI{390}{\mega MAC/\second} @ \SI{203}{\milli \watt}, running at \SI{550}{\mega\hertz}.
The GAP8, at the maximum frequency of CL@\SI{175}{\mega\hertz}, achieves \SI{1.44}{\giga MAC/\second} @ \SI{102}{\milli \watt} and \SI{1.14}{\giga MAC/\second} @ \SI{119}{\milli \watt} with NEMO/DORY and GAP\textit{flow}, respectively.
This analysis shows that, whether using the GAP\textit{flow} or the NEMO/DORY pipeline, the GAP8 outperforms the Cortex-M7-based MCU with CMSIS-NN by more than $1.7\times$ in power consumption and by more than $3\times$ in throughput.
The throughput is of high importance because it has a significant impact on the navigation capabilities, as showed and discussed in Section~\ref{sec:in_field}.

\subsubsection{Power break-down} \label{sec:power_breakdown}

This section analyzes the power-breakdown of the entire nano-UAV running both PULP-Dronet V2 pipelines -- i.e., NEMO/DORY and GAP\textit{flow}.
We frame this investigation also considering -- for each PULP-Dronet V2 version -- the two GAP8's operating points introduced in Section~\ref{sec:power}, called \textit{most energy efficient} and \textit{maximum performance}, respectively running at FC@\SI{50}{\mega\hertz} CL@\SI{100}{\mega\hertz}, and FC@\SI{250}{\mega\hertz} CL@\SI{175}{\mega\hertz}.

The analysis in Figure~\ref{fig:power_breakdown} refers to three main parts: \textit{i}) the nano-drone's motors, \textit{ii}) its basic electronics running the stock flight controller, and \textit{iii}) the AI-deck executing our visual-workloads.
The electronics slice accounts for both flight controller MCUs (i.e., STM32 and nRF51) and all the basic platform's sensors (e.g., IMU, barometer).
Additionally, for all four configurations, we also report a break-down zoom-in on the three main AI-deck's components: \textit{i}) the GAP8 SoC, \textit{ii}) the off-chip DRAM, and \textit{iii}) the ULP camera.
The reader should note that the DRAM is considered active at full speed only for the time required to copy the CNN's parameter from L3 to L2 and otherwise turned off.
Similarly, Flash memory is not considered in this power break-down evaluation, as it is necessary only for the system initialization (i.e., data movement from Flash to DRAM) and then never again used during the drone's mission.

Figure~\ref{fig:power_breakdown} shows how the four motors consume the vast majority of the total power budget, i.e., \SI{7.3}{\watt}, while the rest of the drone's electronics accounts for \SI{277}{\milli\watt} in all four configurations, as they are always kept at the same operative conditions.
Conversely, the power consumption for the AI-deck changes depending on both exploration parameters, but it is never higher than 1.64\%, resulting in a motors' power consumption between 94.8\% and 96.0\% of the total budget.

\begin{figure*} [t]
 \includegraphics[width=\textwidth]{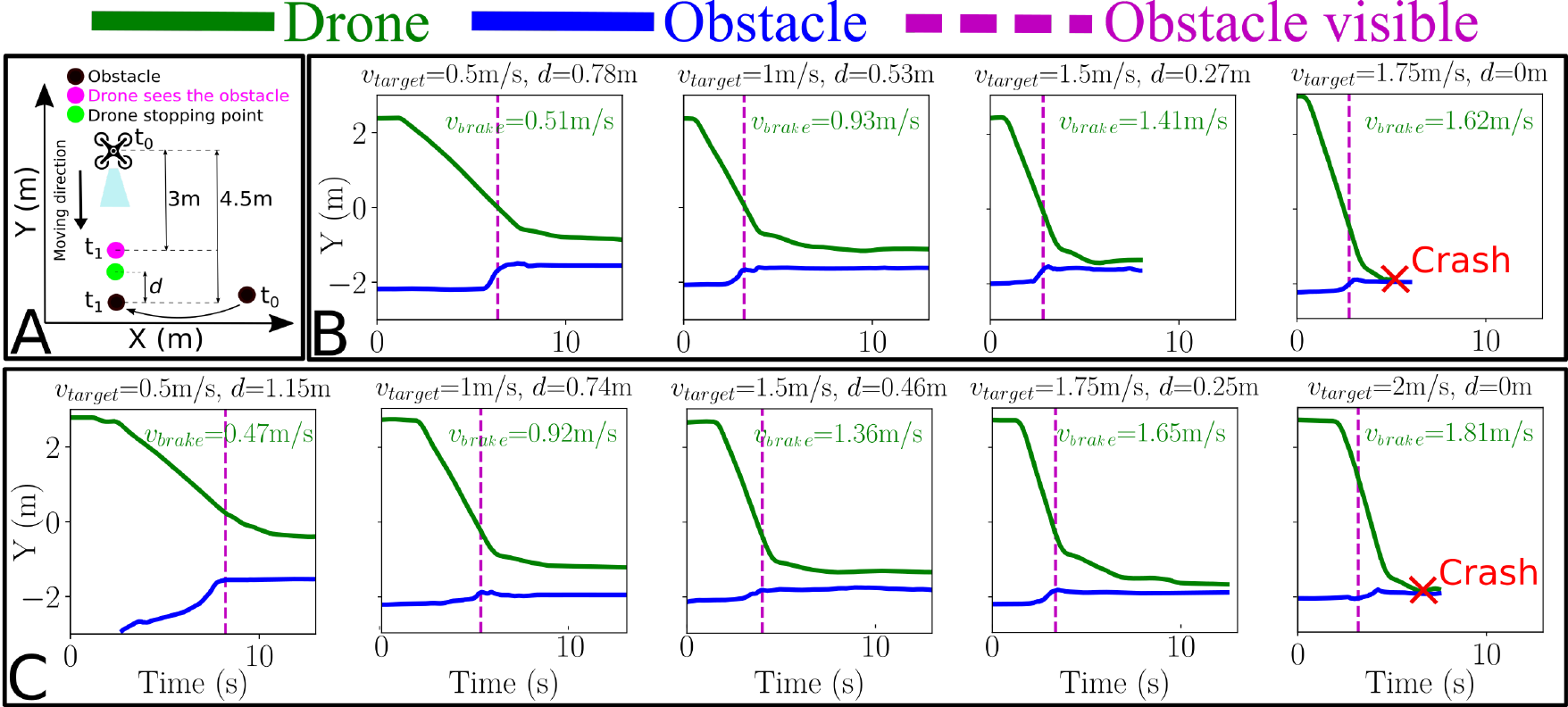}
 \caption{A) Experimental setup for the \textit{obstacle avoidance} evaluation. In-field tests are carried sweeping $v_{target}$, for both the most energy efficient (B) and the maximum performance (C) SoC's configurations.}
 \label{fig:obstacle_avoidance}
\end{figure*}

Focusing on the comparison between the two PULP-Dronet V2 versions, we can identify two main behaviors: \textit{i}) the version developed using the NEMO/DORY framework always shows higher average power consumption for the DRAM and \textit{ii}) the GAP\textit{flow}-based version always has a marginally higher average power consumption for the GAP8 computation.
In the most energy efficient configuration, the NEMO/DORY implementation accounts for the 0.63\% of the system's power consumption (Figure~\ref{fig:power_breakdown}-A), while GAP\textit{flow} accounts for the 0.67\% (Figure~\ref{fig:power_breakdown}-C). 
In fact, as shown in Section~\ref{sec:CL_FC_power_consumption}, the GAP\textit{flow}-based implementation brings, on average, to a slightly higher power consumption compared to NEMO/DORY one, \SI{39}{\milli\watt} and \SI{34}{\milli\watt}, respectively.
Moreover, due to the L2 data pre-loading during the initialization stage, the GAP\textit{flow} version accesses the DRAM fewer times than its counterpart, resulting in a lower DRAM power consumption w.r.t. NEMO/DORY, i.e., \SI{8}{\milli\watt} and \SI{10}{\milli\watt}, respectively.

Moving to a comparison on the SoC's operating points, the max performance configuration gives a slight advantage to the NEMO/DORY version of PULP-Dronet, which makes the AI-deck consuming 1.43\% of the total system's power (Figure~\ref{fig:power_breakdown}-B), while using GAP\textit{flow} the percentage becomes 1.64\% (Figure~\ref{fig:power_breakdown}-D).
This small advantage comes from the fact that the NEMO/DORY version, requiring more L3-L2 data transfer w.r.t. the GAP\textit{flow} version, can benefit more from the increased FC's frequency of the max performance configuration.
This difference results in a minimal reduction of the AI-deck's power consumption for the NEMO/DORY-based version, as much as 0.04\% and 0.21\% compared to the GAP\textit{flow}, for the most energy-efficient configuration and the maximum performance one, respectively.
Therefore, from a practical viewpoint, both PULP-Dronet V2 versions perform with a very similar power envelope when deployed on our nano-drone.
Ultimately, in all four configurations, we can remark how the addition of the AI workload to our quadrotor only accounts for the smallest portion of the power consumption of the entire system, never higher than 1.64\%.
Such a small impact on the whole system's power budget demonstrates the capability of running the PULP-Dronet V2 at the highest performance point, only marginally impacting the quadrotor lifetime.
This enables the possibility to further extend the onboard intelligence with additional tasks (e.g., tracking, detection, localization), aiming at more complex mission objectives.

\subsection{In-field closed-loop evaluation} \label{sec:in_field}

In the following, we perform the in-field evaluation of the navigation capabilities of the PULP-Dronet V2, using the implementation generated by the GAP\textit{flow} framework, and deploying it on a Crazyflie 2.1
nano-drone equipped with an additional AI-deck, as illustrated in Section~\ref{sec:background}.
We focus on four key aspects to assess the performances of our closed-loop nano-UAV: \textit{i}) the obstacle avoidance task; \textit{ii}) the lane following task; \textit{iii}) the longest flight distance in a familiar environment; \textit{iv}) the generalization capability, testing the autonomous navigation in never-seen-before environments.

\subsubsection{Obstacle avoidance task}
One of the two outputs of the Dronet CNN is the probability of collision used to predict a potential obstacle in the path followed by the nano-drone.
In this set of experiments, we assess the drone's robustness in avoiding dynamic obstacles by stressing the closed-loop system with an ad-hoc setup, as shown in Figure~\ref{fig:obstacle_avoidance}-A.
The drone flies a straight trajectory of \SI{4.5}{\meter} where a dynamic obstacle (i.e., $0.7\times$\SI{0.7}{\meter} cardboard sheet) appears at the end of the path, leaving only \SI{1.5}{\meter} for braking and avoiding the collision -- i.e., \textit{braking-space}.
The straight flight is enforced by silencing the steering angle output of the CNN -- i.e., always 0.
We perform and record all experiments in a room equipped with a mm-precise motion capture system @ \SI{50}{\hertz} (i.e., Vicon) to analyze the drone's behavior in post-processing.

We investigate this scenario by sweeping two key parameters: \textit{i}) the drone's \textit{target forward velocity} ($v_{target}$), i.e., a software parameter representing the forward velocity the drone tries to reach if no obstacle is detected, and \textit{ii}) the CNN's inference throughput by means of the two SoC's configurations, introduced in Section~\ref{sec:power}, named \textit{most energy-efficient} and \textit{max performance}.
This evaluation is depicted in Figure~\ref{fig:obstacle_avoidance}-B for most energy-efficient configuration (FC@\SI{50}{\mega\hertz}, CL@\SI{100}{\mega\hertz}) peaking at \SI{8.7}{frame/\second}, and in Figure~\ref{fig:obstacle_avoidance}-C for the maximum performance one (FC@\SI{250}{\mega\hertz}, CL@\SI{175}{\mega\hertz}) up to \SI{12.8}{frame/\second}.

We stress the system with a growing velocity $v_{target}$ that takes the following values: \SI{0.5}{\meter/\second}, \SI{1.0}{\meter/\second}, \SI{1.5}{\meter/\second}, and \SI{1.75}{\meter/\second} in Figure~\ref{fig:obstacle_avoidance}-B, and \SI{0.5}{\meter/\second}, \SI{1.0}{\meter/\second}, \SI{1.5}{\meter/\second}, \SI{1.75}{\meter/\second}, and \SI{2.0}{\meter/\second} in Figure~\ref{fig:obstacle_avoidance}-C.
As the $v_{target}$ is a software parameter, we also report, for each test, the actual peak velocity when the drone begins braking ($v_{brake}$) recorded with the Vicon.
We stop this incremental procedure once we reach the limit for which the nano-drone can not prevent the collision anymore.
Additionally, in all plots, we highlight a dashed vertical line marking the time when the moving obstacle appeared in the drone's view.

The system proves to be fully working up to $v_{brake}=\SI{1.41}{\meter/\second}$ and $v_{brake}=\SI{1.65}{\meter/\second}$, in the most energy-efficient configuration and the maximum performance one, respectively.
We can notice how the maximum performance configuration provides a similar safety distance ($d=\sim\SI{0.25}{\meter}$) w.r.t. its counterpart, despite their different $v_{brake}$, thanks to the higher inference throughput (i.e., \SI{12.8}{frame/\second} vs. \SI{8.7}{frame/\second}).
Lastly, referring to the PULP-Dronet baseline~\cite{palossi2019DCOSS}; for the same FC@\SI{50}{\mega\hertz}, CL@\SI{100}{\mega\hertz} configuration, we score a 25.3\% higher $v_{brake}$/\textit{braking-space} ratio, confirming not only the successful deployment of our PULP-Dronet V2 on the COTS Crazyflie nano-drone but also increased promptness of the system.

\subsubsection{Lane detection task}

\begin{figure*}[t]
\centering
\includegraphics[width=1\linewidth]{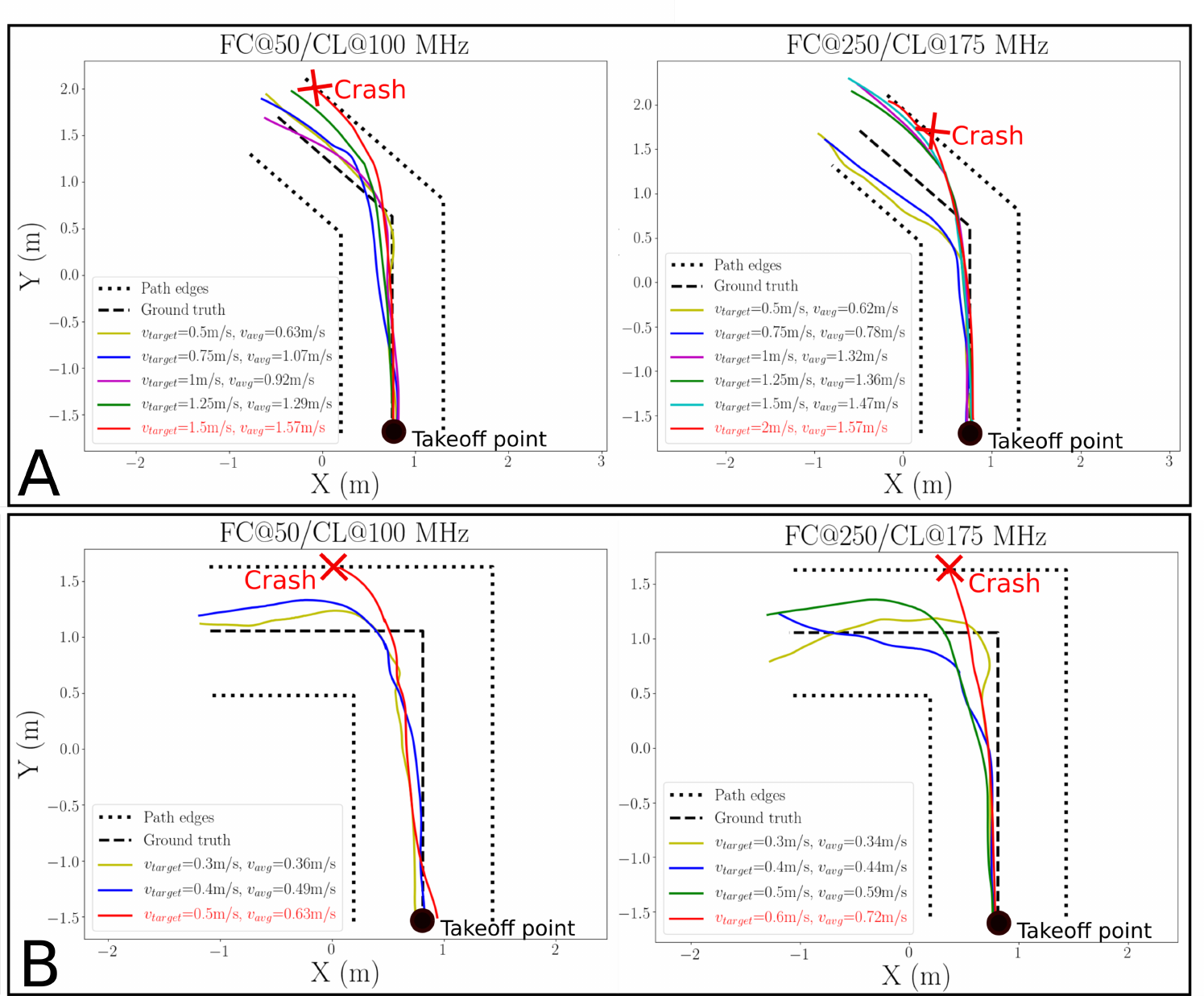}
\caption{Lane detection task evaluation. We assess the CNN's capability of predicting the correct (ground truth) steering angle in a scenario featuring a left-side turn -- \SI{45}{\degree} (A) \SI{90}{\degree} (B) -- at the center of the path. We sweep the forward target velocity ($v_{target}$) identifying the limit of our system -- in red failing configurations.}
\label{fig:steering}
\end{figure*}

In the following experiments, we assess the PULP-Dronet V2 capability to predict the correct steering angle under two controlled curvature scenarios: a smooth turn of \SI{45}{\degree}, and a more challenging \SI{90}{\degree} (i.e., sharp turn).
The setup consists of a path \SI{4.5}{\meter} long and \SI{1.3}{\meter} wide with a left-side turn in the middle, as depicted in Figure~\ref{fig:steering}-A/B with the dotted lines indicating the boundaries of the path.
We explore two parameters \textit{i}) the target forward velocity ($v_{target}$) and \textit{ii}) the CNN's throughput, like in the previous obstacle avoidance experiments, utilizing the two SoC's configurations introduced in Section~\ref{sec:power}.
The software parameter $v_{target}$ is swept with a granularity of \SI{0.25}{\meter/\second} for the \SI{45}{\degree} case, and a smaller \SI{0.1}{\meter/\second} growing-step for the \SI{90}{\degree} setup, due to its higher complexity.

Starting from $v_{target}=\SI{0.5}{\meter/\second}$ for the \SI{45}{\degree} setup (Figure~\ref{fig:steering}-A) and $v_{target}=\SI{0.3}{\meter/\second}$ for the \SI{90}{\degree} one (Figure~\ref{fig:steering}-B), we keep increasing the $v_{target}$ until the system reaches its limit (i.e., collision).
This exploration defines the actual maximum average velocity ($v_{avg}$) -- including the initial acceleration phase -- for which the nano-drone can complete this lane detection task, resulting in:
\begin{itemize}
 \item scenario \SI{45}{\degree}, configuration FC @ \SI{50}{\mega\hertz} CL @ \SI{100}{\mega\hertz}: maximum $v_{avg}=\SI{1.29}{\meter/\second}$;
 \item scenario \SI{45}{\degree}, configuration FC @ \SI{250}{\mega\hertz} CL @ \SI{175}{\mega\hertz}: maximum $v_{avg}=\SI{1.47}{\meter/\second}$;
 \item scenario \SI{90}{\degree}, configuration FC @ \SI{50}{\mega\hertz} CL @ \SI{100}{\mega\hertz}: maximum $v_{avg}=\SI{0.49}{\meter/\second}$;
 \item scenario \SI{90}{\degree}, configuration FC @ \SI{250}{\mega\hertz} CL @ \SI{175}{\mega\hertz}: maximum $v_{avg}=\SI{0.59}{\meter/\second}$;
\end{itemize}
The increased throughput of the \textit{maximum performance} configuration (\SI{12.8}{frame/\second} vs. \SI{8.7}{frame/\second}), enables higher flight speed in both testing scenarios.
As expected, the higher complexity of the \SI{90}{\degree} scenario is confirmed by a lower maximum $v_{avg}$ compared to the \SI{45}{\degree} counterpart.

\begin{table}
\label{tab:RMSE}
\caption{Steering angle RMSE and actual average velocity.}
\centering
\footnotesize
\renewcommand{\arraystretch}{1.1}% for the vertical padding of tables
\begin{tabular}{|c|c|c|cccc|} 
% \hline
% \multicolumn{7}{|c|}{\textbf{Steering angle RMSE \& actual average velocity}} \\ 
\hline
\multicolumn{1}{|c}{} & & & \multicolumn{4}{c|}{SoC configuration} \\ 
\cline{4-7}
\multicolumn{1}{|c}{} & & & \multicolumn{2}{c|}{\begin{tabular}[c]{@{}c@{}}FC@50/CL@100 \\MHz\end{tabular}} & \multicolumn{2}{c|}{\begin{tabular}[c]{@{}c@{}}FC@250/CL@175~\\MHz\end{tabular}} \\ 
\cline{4-7}
\multicolumn{1}{|c}{} & & \begin{tabular}[c]{@{}c@{}}$v_{target}$\\{[}m/s]\end{tabular} & \multicolumn{1}{c|}{\begin{tabular}[c]{@{}c@{}}RMSE\\~{[}m]\end{tabular}} & \multicolumn{1}{c|}{\begin{tabular}[c]{@{}c@{}}$v_{avg}$\\{[}m/s]\end{tabular}} & \multicolumn{1}{c|}{\begin{tabular}[c]{@{}c@{}}RMSE \\{[}m]\end{tabular}} & \begin{tabular}[c]{@{}c@{}}$v_{avg}$\\{[}m/s]\end{tabular} \\ 
\hline
\multirow{8}{*}{\rotcell{\textbf{Turn curvature}}} & \multirow{5}{*}{45°} & 0.5 & \textbf{0.07} & 0.63 & 0.26 & 0.62 \\
 & & 0.75 & 0.12 & 1.07 & 0.21 & 0.78 \\
 & & 1 & \textbf{0.07} & 0.92 & 0.22 & 1.32 \\
 & & 1.25 & 0.17 & \textbf{1.29} & \textbf{0.20} & 1.36 \\
 & & 1.5 & collision & 1.57 & 0.23 & \textbf{1.47} \\ 
\cline{2-7}
 & \multirow{3}{*}{90°} & 0.3 & \textbf{0.14} & 0.36 & \textbf{0.11} & 0.34 \\
 & & 0.4 & 0.17 & \textbf{0.49} & 0.16 & 0.44 \\
 & & 0.5 & collision & 0.63 & 0.20 & \textbf{0.59} \\
\hline
\end{tabular}
\end{table}

The 2D trajectories of all tests are reported in Figure~\ref{fig:steering}, where we define as \textit{ground truth} (dashed lines) the trajectory an ideal nano-drone would follow, keeping the path center for the entire flight.
Comparing the actual trajectories with the ground truth one, in Table~\ref{tab:RMSE}, we report the root mean squared error (RMSE) for all tests.
For both scenarios, we can see a general trend where the higher the actual velocity is ($v_{avg}$) the more the RMSE grows, ranging between 0.07-\SI{0.26}{\meter} and 0.11-\SI{0.20}{\meter}, for the \SI{45}{\degree} and \SI{90}{\degree} case, respectively.

A second interesting trend can be seen in the variation of the RMSE between the two SoC configurations of the \SI{45}{\degree} scenario.
At a first look, it seems that a higher throughput penalizes the system's capability, increasing the RMSE.
However, by looking at the drone's trajectories in Figure~\ref{fig:steering}-A (right plot), it is clear how the successful tests are clustered into two groups: 
\begin{itemize}
 \item tests $v_{target}$ 0.5 and 0.75 (yellow and blue curves) tend to follow a shorter path trajectory;
 \item all the other curves exhibit a right-hand drive policy fostered by the dataset labeled with steering angles (i.e., Udacity samples are collected in the US).
\end{itemize}
In both cases, the ultimate trajectory is slightly away from the ideal central ground truth, increasing the RMSE but still accomplishing the mission.

\subsubsection{Longest flight distance}

In this set of experiments, we want to assess our closed-loop system's autonomous navigation capability in a free-flight mission, exploring a ``friendly'' environment.
For this purpose, we select as mission field the same \SI{110}{\meter}-long corridor (U-shape) used for collecting part (16\%) of the Himax dataset images.
The mission field presents only static obstacles (e.g., walls, doors, and furniture), where we perform 25 tests, sweeping the $v_{target}$ parameter.
We employ a growing-step of \SI{0.5}{\meter/\second}, from $v_{target}=\SI{0.5}{\meter/\second}$ to \SI{2.5}{\meter/\second}, testing each configuration 5 times.
All these experiments are made by selecting the maximum performance SoC's configuration (FC@\SI{250}{\mega\hertz}, CL@\SI{175}{\mega\hertz}), able to deliver up to \SI{12.8}{frame/\second} inference.

\begin{table}
\centering
\caption{Evaluation of the flight time and average velocity when the drone flies through a \SI{110}{\meter} long corridor.}
\label{tab:flight-distance}
\begin{tabular}{|c|c|c|c|c|} 
\hline
\begin{tabular}[c]{@{}c@{}}$v\sb{target}$ \\{[}m/s]\end{tabular} & \begin{tabular}[c]{@{}c@{}}$v\sb{avg}$ \\{[}m/s]\end{tabular} & \begin{tabular}[c]{@{}c@{}}Average\\time [s]\end{tabular} & \begin{tabular}[c]{@{}c@{}}Distance \\{[}m]\end{tabular} & \begin{tabular}[c]{@{}c@{}}Success \\rate\end{tabular} \\ 
\hline
0.5 & 0.51 & 216 & 110 & \textbf{5/5} \\ 
\hline
1 & 0.98 & 112 & 110 & \textbf{5/5} \\ 
\hline
1.5 & 1.72 & 64 & 110 & \textbf{5/5} \\ 
\hline
2 & 1.96 & 56 & 110 & 4/5 \\ 
\hline
2.5 & \textbf{2.29} & \textbf{48} & 110 & 1/5 \\
\hline
\end{tabular}
\end{table}

\begin{figure*}[t]
 \includegraphics[width=\textwidth]{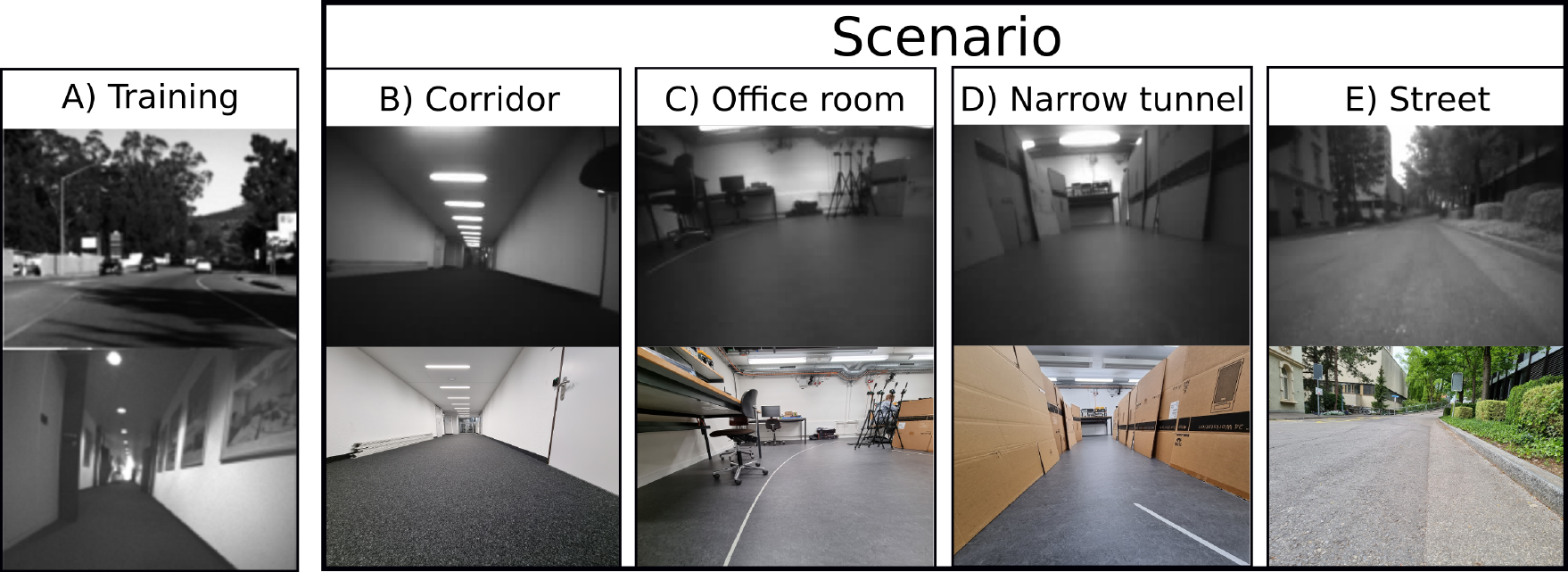}
 \caption{Samples of: A) the training images (both Udacity and Himax dataset); B) working-place corridor; C) office room with furniture; D) narrow pipeline-like tunnel; E) public street.}
 \label{fig:scenarios}
\end{figure*}

In Table~\ref{tab:flight-distance}, we summarize all the experiments for each $v_{target}$ configuration, reporting the average flight time over the successful runs and highlighting in bold the peak performances.
We achieve the maximum success-rate (5 success out of 5 tests, per configuration), with an actual mean flight velocity ($v_{avg}$) from 0.51 to \SI{1.72}{\meter/\second}.
Increasing the $v_{avg}$ to \SI{1.96}{\meter/\second}, lowers the the success-rate to 80\%, defining the performance upper bound of our system, as increasing even further the $v_{avg}$ to \SI{2.29}{\meter/\second} the success-rate quickly drops to 20\%.
With such high velocity also comes an increased acceleration the drone applies to reach the desired $v_{target}$, resulting in a high positive pitch.
Therefore, the camera mostly captures the floor, which turns in a minimal time to react to the obstacles, flying at high speed. 
These results prove a superior performance compared to PULP-Dronet V1, which for the same flown distance reports an average velocity of \SI{0.5}{\meter/\second}.
While PULP-Dronet V1 used a linear mapping between the probability of collision and the forward velocity, we extended this mechanism with a quadratic relation. 
This results in a more significant reduction in the forward velocity while approaching an obstacle, which gives the drone more time to steer when flying at high velocities (prior to steering)~\footnote{As supplementary material, we make available video footage of one run, with $v_{target}=\SI{2.0}{\meter/\second}$, available at \url{https://youtu.be/41IwjAXmFQ0}.}.
The main limitation of this experiment is that the corridor is ``known'' to PULP-Dronet V2 as the training set contains images of the corridor: in the next section, we explore the network's capability to generalize to never-seen-before testing environments.

\subsubsection{Generalization capability}

As the last part of our in-field evaluation, we present a set of functional experiments aiming at demonstrating the robustness of the PULP-Dronet V2 CNN, testing the closed-loop system in different deployment scenarios.
As reported in Table~\ref{tab:scenarios}, and showcased in Figure~\ref{fig:scenarios}, we explore four different application scenarios, namely:
\begin{enumerate}
 \item \textbf{corridor:} a working-place corridor that is not part of the Himax dataset;
 \item \textbf{office room:} a room with tables and lab equipment;
 \item \textbf{narrow tunnel:} a narrow pipeline-like tunnel ($\sim$\SI{5}{\metre} long and $\sim$\SI{1.2}{\metre} wide) made of cardboard;
 \item \textbf{street:} a public street, with road signs and cars.
\end{enumerate}
Orthogonal to the four scenarios, we also consider a second testing criterion: the presence/absence of obstacles in the path the nano-drone should follow.
For the cases \textit{corridor}, \textit{office room}, and \textit{street}, the obstacle is represented by a person that can be either standing still in the center of the path (i.e., static obstacle) or moving and crossing the trajectory of the drone or stopping in front of it (i.e., dynamic obstacle). 
Regardless of the type of obstacle, the goal is to adjust the moving direction, avoiding the obstacle.
Instead, for the \textit{narrow tunnel} case, we employ a small cardboard panel as an obstacle, still differentiating between a static and dynamic configuration.
All these experiments refer to a setup with a drone's mean target velocity $v_{target}=\SI{0.5}{\metre / \second}$. 

Considering the training datasets shown in Figure~\ref{fig:scenarios}-A, the selected deployment fields introduce a significant difference between the visual cues present in the in-field images and those the CNN has been trained with.
The main two sources of difference between the training and deployment can be ascribed to \textit{i}) environmental conditions (e.g., absence of road lane signs in the street) \textit{ii}) photometric/geometric differences between the cameras used to acquire the vast majority of the training dataset and the one available on the mission drone (e.g., field-of-view and resolution).

\begin{table}
\centering
\caption{In-field evaluation scenarios over multiple runs with and without obstacles (mean flight time : success-rate).}
\label{tab:scenarios}
\begin{tabular}{|c|l|c|c|c|} 
\hline
\multirow{6}{*}{\rotcell{\textbf{Scenario}}} & & \multicolumn{3}{|c|}{\textbf{Obstacle}} \\ 
\cline{3-5}
& & None & Static & Dynamic \\ 
\cline{2-5}
& 1. Corridor & \SI{144}{\second} : 6/6 & \SI{137}{\second} : 3/6 & \SI{63}{\second} : 4/6 \\ 
\cline{2-5}
& 2. Office room & \SI{86}{\second} : 5/6 & \SI{78}{\second} : 4/6 & \SI{67}{\second} : 5/6 \\ 
\cline{2-5}
& 3. Narrow tunnel & \SI{12}{\second} : 5/6 & \SI{0}{\second} : 0/6 & \SI{19}{\second} : 4/6 \\ 
\cline{2-5}
& 4. Street & \SI{171}{\second} : 6/6 & \SI{148}{\second} : 6/6 & \SI{148}{\second} : 6/6 \\
\hline
\end{tabular}
\end{table}

In Table~\ref{tab:scenarios}, we report the results in terms of mean flight time (across multiple runs) and success-rate, for each case.
For scenarios 1, 2, and 4, we consider the test successful if the drone follows the path, with no crashes, for at least \SI{60}{\second}.
Instead, for scenario 3, we define as success criteria the capability of the nano-drone to complete the exploration of the entire narrow tunnel.
Among all successful cases, the mean flight time spans from \SI{12}{\second} to \SI{171}{\second}, for scenario 3 with no obstacles and scenario 4 with no obstacles, respectively.

The PULP-Dronet V2 sample application shows a high success rate for all configurations except for the \textit{narrow tunnel} with static obstacles, in which case the nano-drone gets stuck in the obstacle proximity, slowly drifting towards it, until it suddenly crashes.
As introduced in Section~\ref{sec:background}, the training datasets have disjoint labels, with the Udacity set providing only steering labels and both Bicycle and Himax sets coupled with only collision ones.
Therefore, the actual tasks learned by the CNN can be defined as \textit{``predict the presence of obstacles''} and \textit{``predict the steering to follow the lane''}, but not explicitly \textit{``predict the steering to prevent a collision''}.
This limit of the Dronet CNN (all versions) could be mitigated in the future by either introducing a new training dataset providing both labels for each image sample or introducing an additional level of intelligence between the CNN and the low-level control. 

However, this group of tests aims at assessing the generalization capability of the PULP-Dronet V2 baseline model without introducing any additional deep learning technique, such as deep domain adaptation~\cite{sun2016deep}, dataset augmentation~\cite{palossi2021fully} or continual learning~\cite{ren2021tinyol}, as they would be out of the scope of this work.
All the nine configurations presented in this evaluation are provided of video footage of the experiments available at \url{https://youtu.be/Cd9GyTl6tHI}.

\section{Conclusion}\label{sec:conclusion}

The recent progress in deep learning research opens new possibilities for using vision-based end-to-end neural networks to enable nano-drone autonomous navigation.
The MCUs found in nano-drones have limited memory and computational resources, and therefore they are unable to run complex CNN models in their original form.
However, the available solutions for complexity reduction of the CNNs used to facilitate navigation mainly involve hand-crafted modifications and typically require multiple iterations.
This paper fills this gap by analyzing and integrating tools and methodologies that automate this optimize-and-deploy process, automating fine-grained hardware-aware tuning of the CNN.
We perform an extensive experimental evaluation of the proposed flow, using a SoA CNN for autonomous nano-drone navigation~\cite{palossi2019IOTJ}, achieving $\sim3-$\SI{4}{\milli \joule/frame} inference and reducing the memory requirements by $2\times$ and improving the throughput by $1.6\times$ while preserving the same $\sim90\%$ classification accuracy of the original implementation.
Furthermore, we perform an in-field evaluation of the navigation capabilities of a nano-drone, considering the collision avoidance, steering capabilities, maximum flown distance, and generalization in never-seen-before environments.
We record a maximum indoor flight distance of \SI{110}{\meter} and an average velocity of \SI{1.96}{\meter/\second}, $4\times$ higher than our PULP-Dronet baseline.
We foster the research community releasing as open-source our code and models: \url{https://github.com/pulp-platform/pulp-dronet}.

\section*{Acknowledgments}
This work has been partially funded by the ARR Center of the UAE Technology Innovation Institute (TII).
The authors would like to thank C. Cioflan, P. Mayer, and M. Nikolov for their assistance in recording the supplementary videos.


\newpage
\begin{IEEEbiography} [{\includegraphics[width=1in,height=1.25in,clip,keepaspectratio]{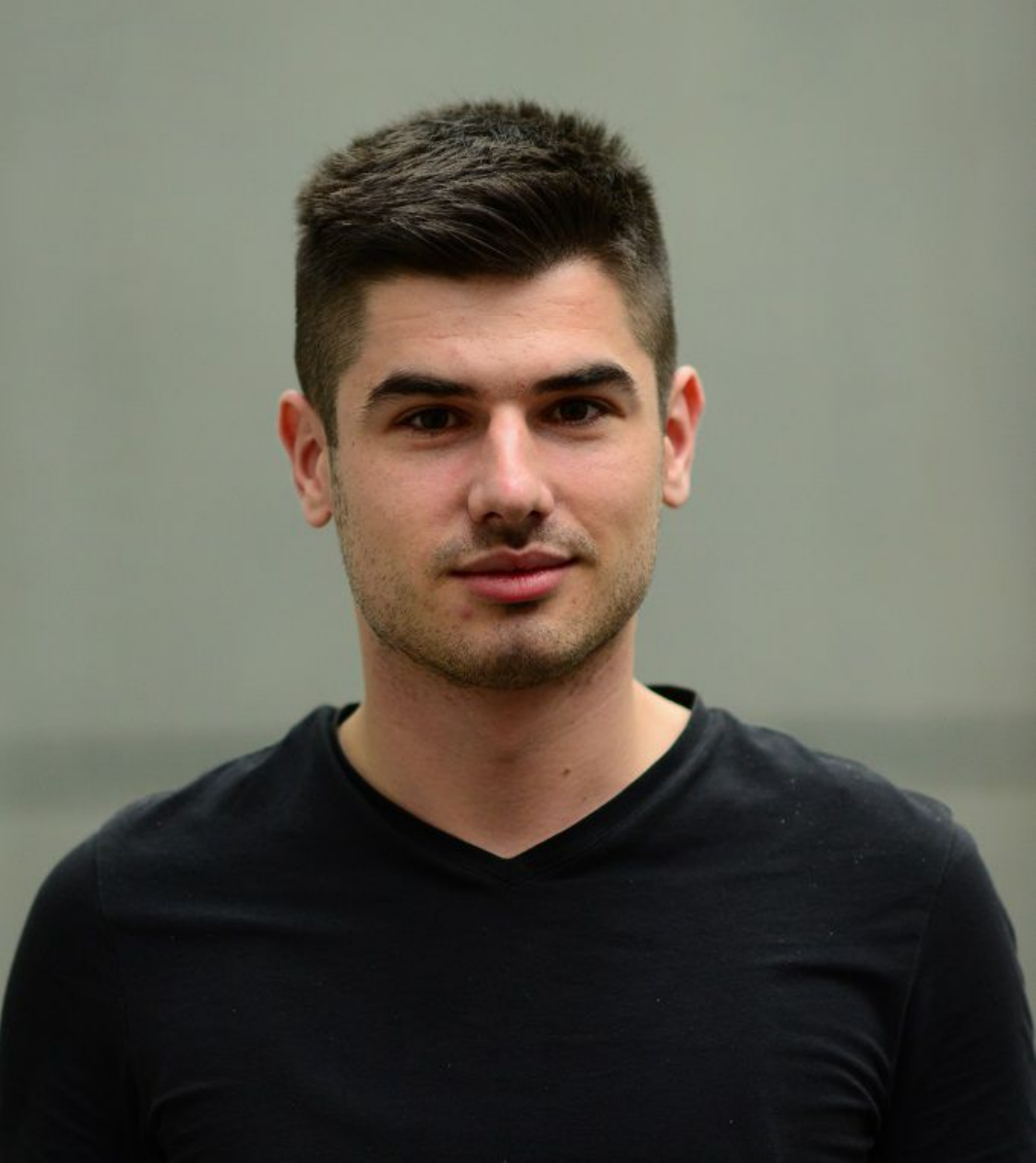}}]
{Vlad Niculescu}
received the Master's degree in Robotics, Systems, and Control from the ETH Zürich, in 2019. He is currently pursuing the Ph.D. in Electrical Engineering within the Integrated Systems Laboratory in ETH Zürich. During the Bachelor and Master period, he competed in more than ten international student competitions, and he was the electrical lead of the student project Swissloop, which won second place and the innovation award in the SpaceX Hyperloop Pod Competition 2019.
His research is now focused on developing localization and autonomous navigation algorithms that target ultra-low-power platforms which can operate onboard nano-drones.
\end{IEEEbiography}
\begin{IEEEbiography}[{\includegraphics[width=1in,height=1.25in,clip,keepaspectratio]{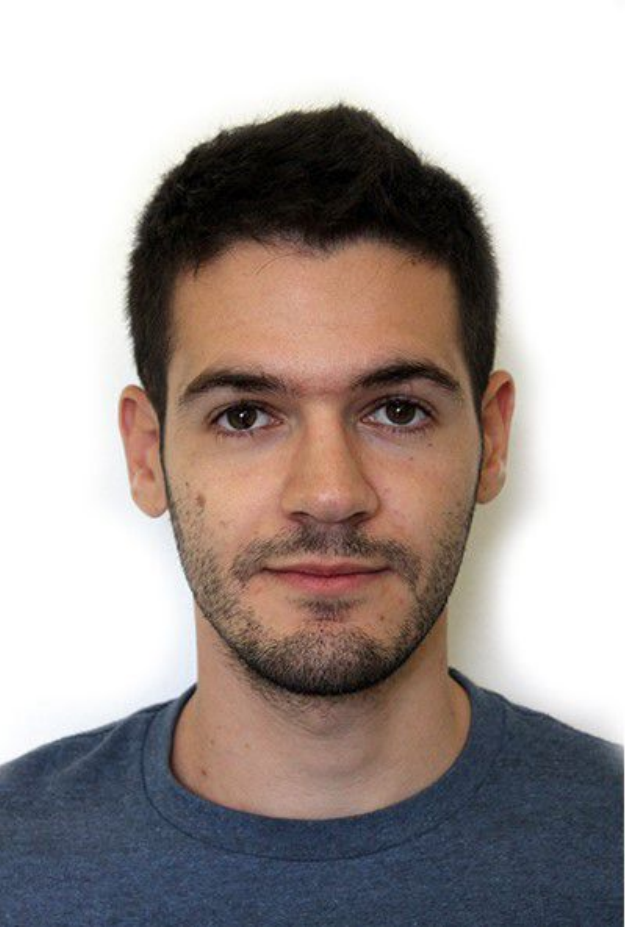}}]
{Lorenzo Lamberti}
graduated with honors in both Bachelor's (2016) and Master's (2019) degrees at the University of Bologna, Italy, where he is now pursuing a Ph.D. in Electronic Engineering in the Energy-Efficient Embedded Systems Laboratory. Previously, he has been an intern at the Fermi National Accelerator Laboratory of Chicago, US, and at the Datalogic's Artificial Intelligence Laboratory in Pasadena, US. His research is currently targeted at autonomous navigation for nano-size unmanned aerial vehicles. In particular, he focuses on neural architecture search, training, and in-field deployment of Deep Neural Network-based navigation tasks on low-power MCUs.
\end{IEEEbiography}
\begin{IEEEbiography}[{\includegraphics[width=1in,height=1.25in,clip,keepaspectratio]{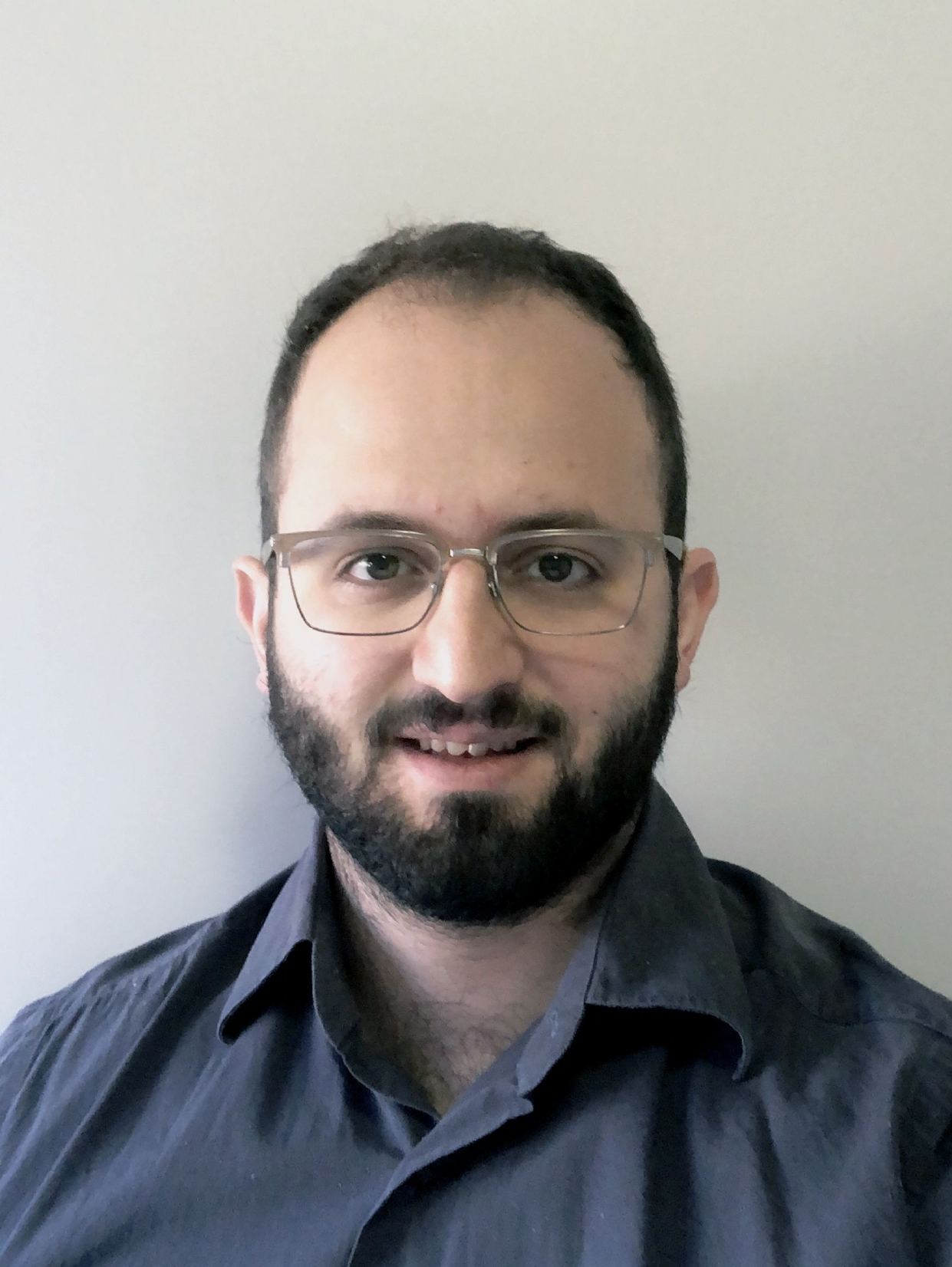}}]%
{Francesco Conti}
received the Ph.D. degree in electronic engineering from the University of Bologna, Italy, in 2016. He is currently an Assistant Professor in the DEI Department of the University of Bologna. From 2016 to 2020, he held a research grant in the DEI department of University of Bologna and a position as postdoctoral researcher at the Integrated Systems Laboratory of ETH Zurich in the Digital Systems group. His research focuses on the development of deep learning based intelligence on top of ultra-low power, ultra-energy efficient programmable Systems-on-Chip. His research work has resulted in more than 60 publications in international conferences and journals and has been awarded several times, including the 2020 IEEE TCAS-I Darlington Best Paper Award.
\end{IEEEbiography}
\begin{IEEEbiography} [{\includegraphics[width=1in,height=1.25in,clip,keepaspectratio]{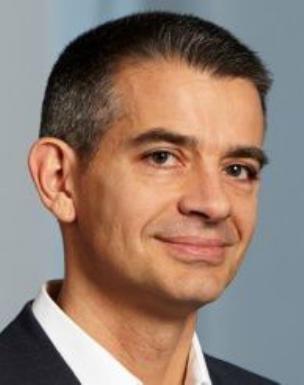}}]%
{Luca Benini}
(Fellow, IEEE) received the Ph.D. degree in electrical engineering from Stanford University, Stanford, CA, USA, in 1997. He has served as the Chief Architect of the Platform2012/STHORM Project with STMicroelectronics, Grenoble, France, from 2009 to 2013. Currently, he holds the Chair of Digital Circuits and Systems at ETH Zurich, Zurich, Switzerland, and is a Full Professor at the University of Bologna, Bologna, Italy. He has published more than 1000 peer-reviewed articles and five books. His current research interest includes energy-efficient computing systems' design from embedded to high performance. Dr. Benini is a fellow of the ACM and a member of the Academia Europaea. He was a recipient of the 2016 IEEE CAS Mac Van Valkenburg Award and the 2019 IEEE TCAD Donald O. Pederson Best Article Award.
\end{IEEEbiography}
\begin{IEEEbiography}[{\includegraphics[width=1in,height=1.25in,clip,keepaspectratio]{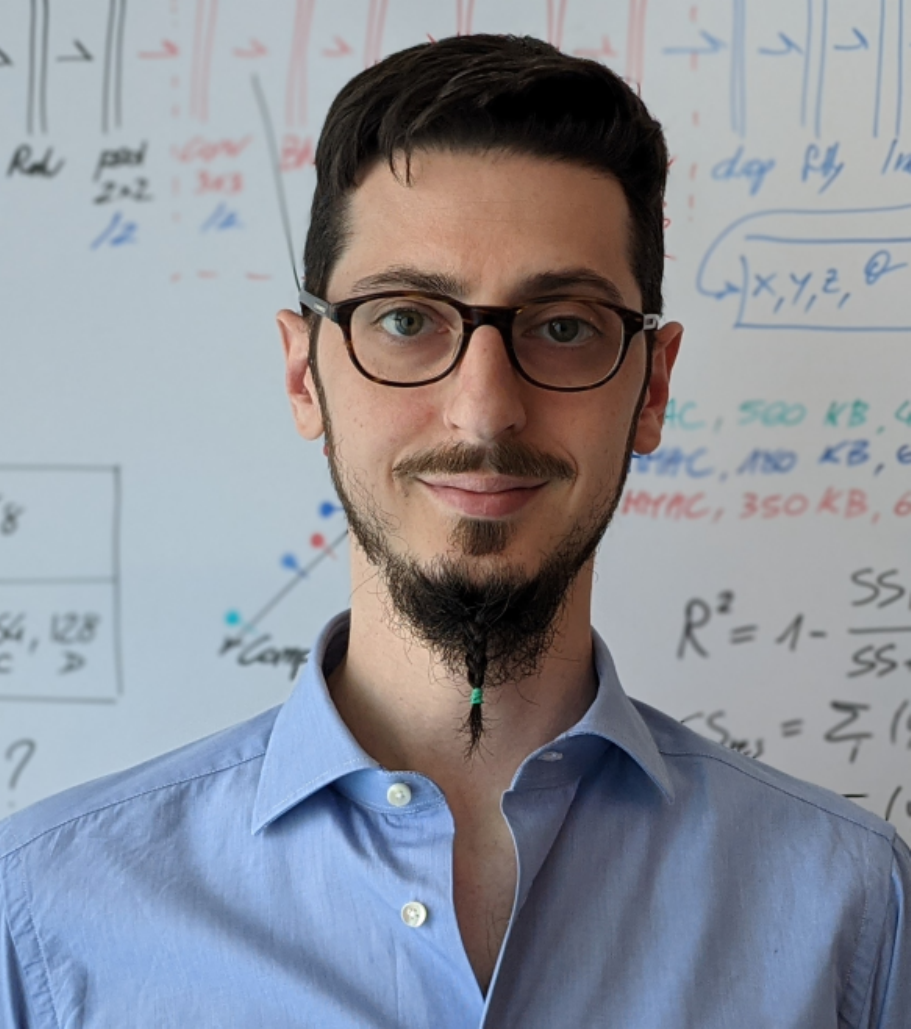}}]%
{Daniele Palossi}
received the Ph.D. in Information Technology and Electrical Engineering from the ETH Zürich, in 2019. He is currently a Postdoctoral Researcher at the Dalle Molle Institute for Artificial Intelligence (IDSIA), USI-SUPSI, Lugano, Switzerland, and at the Integrated Systems Laboratory (IIS), ETH Zürich, Zürich, Switzerland. His research focuses on the embedded domain with special emphasis on energy-efficient ultra-low-power platforms, algorithms for autonomous navigation, and resource-constrained small-sized cyber-physical systems. His work has resulted in 20+ publications in international conferences and journals. Dr. Palossi was a recipient of the Swiss National Science Foundation (SNSF) Spark Grant and the 2nd prize at the Design Contest held at the ACM/IEEE ISLPED'19.
\end{IEEEbiography}

\end{document}